\documentclass[12pt,a4paper]{article}
\usepackage[utf8]{inputenc}
\usepackage[T1]{fontenc} 
\usepackage[english]{babel}
\usepackage[top=3.0cm, bottom=3.5cm, left=3.0cm, right=3.0cm]{geometry}
\usepackage{setspace}
\usepackage{dutchcal}
\usepackage{amssymb}
\usepackage{amsmath}
\usepackage{amsfonts}
\usepackage{mathrsfs}
\usepackage{bm}
\numberwithin{equation}{section}
\allowdisplaybreaks
\usepackage{enumitem}
\usepackage{array}

\usepackage[]{caption} 
\usepackage{hyperref}
\hypersetup{
    colorlinks,%
    citecolor=blue,%
    filecolor=blue,%
    linkcolor=blue,%
   	urlcolor=blue,
   	linktoc=page
}
\usepackage{cite}
\usepackage[toc,page]{appendix}
\usepackage{graphicx}
\usepackage{float}
\usepackage{subfig}
\usepackage{tikz}
\usepackage{xcolor}
\definecolor{DarkGreen}{RGB}{0,128,0}
\newcommand{\p}{\partial}

\newcommand{\ndelta}{\delta\hspace{-0.50em}\slash\hspace{-0.05em} }
\newcommand{\poubelle}[1]{}

\renewcommand{\textbf}[1]{\begingroup\bfseries\mathversion{bold}#1\endgroup}

\begin{document}

\setstretch{1.25}

\setcounter{tocdepth}{2}

\begin{titlepage}

\begin{flushright}\vspace{-3cm}
{\small
\today }\end{flushright}

\begin{center}
\setstretch{1.75}
{\LARGE{\textbf{ The $\Lambda$-BMS$_4$ group of dS$_4$ and \\ new boundary conditions for AdS$_4$
}}}
\end{center}
 \vspace{7mm}
\begin{center} 
\centerline{\large{\bf{Geoffrey Comp\`{e}re\footnote{e-mail: gcompere@ulb.ac.be}, Adrien Fiorucci\footnote{e-mail: afiorucc@ulb.ac.be}, Romain Ruzziconi\footnote{e-mail: rruzzico@ulb.ac.be}}}}

\vspace{2mm}
\normalsize
\bigskip\medskip
\textit{Universit\'{e} Libre de Bruxelles and International Solvay Institutes\\
CP 231, B-1050 Brussels, Belgium\\
\vspace{2mm}
}
\vspace{10mm}

\begin{abstract}
Using the dictionary between Bondi and Fefferman-Graham gauges, we identify the analogues of the Bondi news, Bondi mass and Bondi angular momentum aspects at the boundary of generic asymptotically locally (A)dS$_4$ spacetimes. 
We introduce the $\Lambda$-BMS$_4$ group as the residual symmetry group of the metric in Bondi gauge after boundary gauge fixing. This group consists of infinite-dimensional non-abelian supertranslations and superrotations and it reduces in the asymptotically flat limit to the extended BMS$_4$ group. Furthermore, we present new boundary conditions for asymptotically locally AdS$_4$ spacetimes which admit $\mathbb R$ times the group of area-preserving diffeomorphisms as the asymptotic symmetry group. The boundary conditions amount to fix 2 components of the holographic stress-tensor while allowing 2 components of the boundary metric to fluctuate. They correspond to a deformation of a holographic CFT$_3$ which is coupled to a fluctuating spatial metric of fixed area. \\[0.5cm]

\end{abstract}


\end{center}

\end{titlepage}

\newpage

\tableofcontents

\newpage
\section{Introduction and summary of the results}
\label{sec:intro}

Boundary conditions are required in asymptotically flat and asymptotically anti-de Sitter spacetimes in order to define the asymptotic symmetry group, which determines the conserved surface charges of interest such as the total energy. Several attempts have been made to define boundary conditions at the future boundary $\mathscr I^+$ of de Sitter spacetime, including \cite{Strominger:2001pn,Anninos:2010zf,Anninos:2011jp}. However, as recently emphasized in \cite{Ashtekar:2014zfa,Ashtekar:2015lla}, that enterprise comes with a strong drawback: imposing future boundary conditions amounts to restrict the initial data of a Cauchy slice and, therefore, of the bulk dynamics. Allowing generic initial data and, in particular, generic bulk gravitational waves, prevents imposing any boundary conditions at the future boundary of de Sitter. An interesting question is then how to extend the definition of Bondi mass, Bondi angular momentum and Bondi news in asymptotically locally de Sitter spacetimes, see e.g. \cite{Smalley:1978qd,Abbott:1981ff,Balasubramanian:2001nb,Kastor:2002fu,Bishop:2015kay,Ashtekar:2015ooa,Szabados:2015wqa,Chrusciel:2016oux,Saw:2016isu,Saw:2017amv,Szabados:2018erf,He:2018ikd,Poole:2018koa,Balakrishnan:2019zxm,Mao:2019ahc}. The approach that we will follow here is to gauge fix the gravitational field close to future null infinity in order to isolate the structure that generalizes the BMS$_4$ group \cite{Bondi:1962px,Sachs:1962wk} and its extension \cite{Barnich:2009se,Campiglia:2014yka} to asymptotically locally de Sitter spacetimes. We will also use the dictionary between Bondi gauge and Fefferman-Graham gauge \cite{Starobinsky:1982mr,Fefferman:1985aa} as a guide. 

It is well-known that the gravitational field is entirely determined close to the future boundary $\mathscr I_{\text{dS}}^+$ of asymptotically locally dS$_4$ spacetimes with $\Lambda= 3 \ell^{-2}$ by a 3-dimensional Euclidean boundary metric $g^{(0)}_{ab}$ and a symmetric transverse traceless stress-tensor $T^{ab}$ with respect to this boundary metric up to residual boundary diffeomorphisms and Weyl transformations, see e.g. \cite{Starobinsky:1982mr,Fefferman:1985aa,Skenderis:2002wp,2007arXiv0710.0919F,Papadimitriou:2010as}. Upon further gauge fixing to fixed boundary radial gauge $g^{(0)}_{tt} =\ell^{-2}$, $g^{(0)}_{tA} =0$ and volume $\sqrt{g_{(0)}}$, one can isolate the only dynamical degrees of freedom: the 2 free components of $g^{(0)}_{AB}$ (where $A,B$ span the codimension 2 indices) and the symmetric transverse traceless stress-tensor $T^{ab}$. The residual gauge transformations form a group that we call the $\Lambda$-BMS$_4$ group. It consists of non-abelian supertranslations and superrotations that depend upon arbitrary functions of the coordinates $x^A$. This group exactly reduces to the extended BMS$_4$ group \cite{Campiglia:2014yka} in the asymptotically flat limit.  This is the first result of this paper\footnote{The future boundary $\mathscr I_{\text{dS}}^+$ of dS$_4$ is a three-sphere $S^3$ whose radial slicing singles out two antipodal two-spheres centered at the north and south poles. If boundary conditions are imposed on these 2-spheres as done in \cite{Kelly:2012zc,Ashtekar:2015ooa,Ashtekar:2019khv}, the $\Lambda$-BMS$_4$ group is truncated to a subgroup. However, such future boundary conditions also constrain the initial data and hence we shall not consider them here.}. Due to the presence of the scale $\Lambda$, the structure constants of the BMS$_4$ algebra are generalized and explicitly depend upon the dynamical boundary metric $g_{AB}^{(0)}$. The $\Lambda$-BMS$_4$ algebra is therefore a Lie algebroid \cite{2000math.....12106F,Lyakhovich:2004kr,Barnich:2010eb,Barnich:2010xq}. When $\Lambda \neq 0$, the commutator of supertranslations generates superrotations which are therefore indissociable, contrary to the asymptotically flat limit.

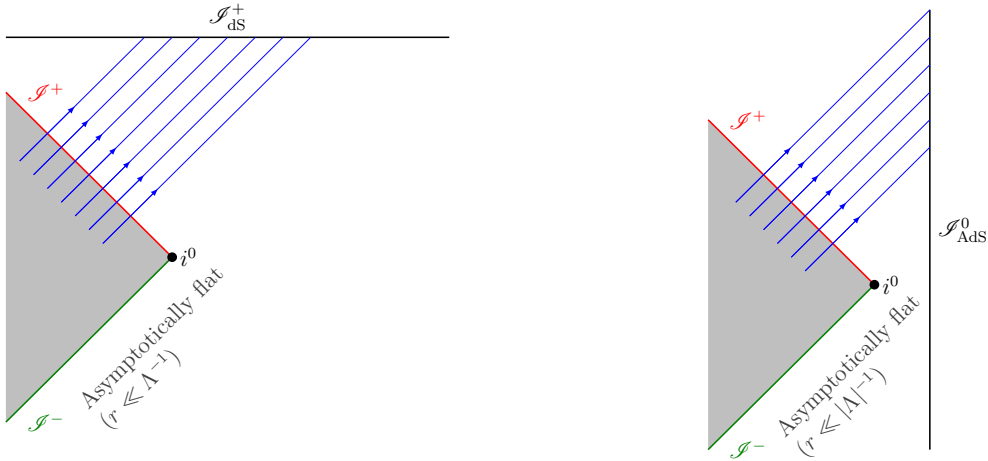
\begin{figure}[!htb]
\centering

\subfloat[][Asymptotically locally dS$_4$ case.]{
	\resizebox{0.5\textwidth}{!}{
	\begin{tikzpicture}
	\draw[white] (-5,-4) -- (-5,5) -- (5,5) -- (5,-4);
	\draw[thick] (-4,4) -- (4,4);
	\draw[] (0,4) node[above]{$\mathscr{I}^+_{\text{dS}}$};
	\fill[black!25] (-4,3) -- (-1,0) -- (-4,-3);
	\draw[red,thick] (-4,3) node[right]{\phantom{..}$\mathscr{I}^+$} -- (-1,0);
	\node[black!75,right,text width=5cm,rotate=45] at (-2.5,-3) {Asymptotically flat\\ ($r \ll \Lambda^{-1}$)};
	\draw[DarkGreen,thick] (-1,0) -- (-4,-3) node[right]{\phantom{..}$\mathscr{I}^-$};
	\node[circle,fill=black,inner sep=0pt,minimum size=5pt] at (-1,0) {};
	\node[right] at (-1,0) {$i^0$};
	\foreach \k in {-3.75,-3.50,...,-2.25} 
	{
	\coordinate (pk) at (\k,-\k-2);
	\coordinate (qk) at (6+2*\k,4);
	\draw[blue] (pk) -- (qk);
	\draw[blue,-latex] (pk) -- ++(1.0,1.0);
	}
	\end{tikzpicture}
	}
}\subfloat[][Asymptotically locally AdS$_4$ case.]{
	\resizebox{0.5\textwidth}{!}{
	\begin{tikzpicture}
	\draw[white] (-3,-4.5) -- (-3,4.5) -- (7,4.5) -- (7,-4.5);
	\draw[thick] (4,-4) -- (4,4);
	\draw[] (4,0) node[right]{$\mathscr{I}^0_{\text{AdS}}$};
	\fill[black!25] (0,2) -- (3,-1) -- (0,-4);
	\draw[red,thick] (0,2) node[right]{\phantom{..}$\mathscr{I}^+$} -- (3,-1);
	\draw[DarkGreen,thick] (3,-1) -- (0,-4) node[right]{\phantom{..}$\mathscr{I}^-$};
	\node[circle,fill=black,inner sep=0pt,minimum size=5pt] at (3,-1) {};
	\node[right] at (3,-1) {$i^0$};
	\node[black!75,right,text width=5cm,rotate=45] at (1.4,-4) {Asymptotically flat \\ ($r \ll |\Lambda|^{-1}$)};
	\foreach \k in {-3.5,-3.25,...,-2.25} 
	{
	\coordinate (pk) at (\k+4,-\k-3);
	\coordinate (qk) at (4,-3-2*\k);
	\draw[blue] (pk) -- (qk);
	\draw[blue,-latex] (pk) -- ++(1.0,1.0);
	}
	\end{tikzpicture}
	}
}
\captionsetup{labelfont=sc,font={small}}
\caption{ 
\setstretch{1.25} On scales $r \ll 1/\vert\Lambda\vert$, the gravitational field can be described without the cosmological constant. One can therefore consider approximately (general) asymptotically flat  regions of any locally asymptotically (A)dS$_4$ spacetime. Lines of constant $u$ (depicted with an arrow in the positive $r$ direction) map the respective boundary $\mathscr I^+_{\text{dS}}$ or $\mathscr I^0_{\text{AdS}}$ to the future null boundary $\mathscr I^+$. The residual $\Lambda$-BMS symmetries are defined in the bulk and admit a smooth flat limit. Since energy flows in the bulk of spacetime, there is no smooth flat limit to $\mathscr I^+$ of the Bondi news and mass defined on $\mathscr I^+_{\text{dS}}$ and $\mathscr I^0_{\text{AdS}}$.}
\label{Fig1}
\end{figure}

After taking into account the counterterm subtraction \cite{Balasubramanian:1999re,deHaro:2000vlm,Skenderis:2002wp,Compere:2008us}, the symplectic structure of Einstein gravity at future null infinity is finite and exactly reduces to
\begin{eqnarray}\label{flux1}
\frac{3}{32 \pi G \ell^4}\int_{\mathscr I_{\text{dS}}^+} d^3 x \sqrt{g_{(0)}}\, \delta J^{AB} \wedge \delta g^{(0)}_{AB}
\end{eqnarray}
where $3(16 \pi G \ell)^{-1} J^{AB}=T^{AB} - \frac{1}{2} g^{AB}_{(0)}T^C_{\,\, C}$ is the traceless part of $T^{AB}$, see \eqref{omegac}-\eqref{Simplectic flux}-\eqref{Symplectic flux after bgf} for the derivation. Hence, the analogue of the Bondi news in asymptotically de Sitter spacetimes is the symplectic pair $(g_{AB}^{(0)},J^{AB})$ which determines the flux of energy leaking through future null infinity. This is the second result of this paper. Accordingly, the quantity $C_{AB}$ that appears in the asymptotic expansion of the metric at the same order as the standard shear is determined in terms of the free data $g_{AB}^{(0)}$ when $\Lambda \neq 0$, as also observed in \cite{Ashtekar:2014zfa,He:2015wfa,Saw:2017amv,He:2018ikd,Poole:2018koa}. While the residual $\Lambda$-BMS$_4$ symmetry extends in the bulk of spacetime and therefore admits a smooth asymptotically flat limit, the fields $(g_{AB}^{(0)},J^{AB})$ are only defined for non-vanishing cosmological constant.  We define the Bondi mass and Bondi angular momentum aspects from the holographic stress-tensor consistently with \cite{Balasubramanian:2001nb}. The evolution of the Bondi mass is determined by the pair $(g_{AB}^{(0)},J^{AB})$. In the asymptotically flat limit, the region $\mathscr I^+_{\text{dS}}$ and the null region $\mathscr I^+$ are distinct and energy flows between the two regions. This is illustrated on Figure \ref{Fig1}. Therefore, the definition of Bondi mass and angular momentum aspects is discontinuous in the limit $\Lambda \rightarrow 0$. The asymptotically flat Bondi mass is $M$ while the Bondi mass for $\text{(A)dS}_4$ $M^{(\Lambda)}$ is given in \eqref{eq:hatM}. 

Upon analytic continuation, asymptotically locally AdS$_4$ spacetimes with $\Lambda = - 3 \ell^{-2}$ are similarly determined close to spatial infinity $\mathscr I^0_{\text{AdS}}$ by a Lorentzian boundary metric $g_{ab}^{(0)}$ and a symmetric transverse traceless stress-tensor $T^{ab}$ up to residual boundary diffeomorphisms and Weyl transformations. The analogous gauge fixing consists in imposing temporal gauge $g^{(0)}_{tt} =-\ell^{-2}$, $g^{(0)}_{tA} =0$ and a fixed volume $\sqrt{-g_{(0)}}$. This is a choice of Dirichlet boundary condition on the boundary metric. The symplectic flux at spatial infinity $\mathscr I^0_{\text{AdS}}$ is then given by \cite{Balasubramanian:1999re,deHaro:2000vlm,Compere:2008us}
\begin{eqnarray}
\frac{3}{32 \pi G\ell^4}\int_{\mathscr I^0_{\text{AdS}}} d^3 x \sqrt{-g_{(0)}}\, \delta J^{AB} \wedge \delta g^{(0)}_{AB},
\end{eqnarray}
which is the analytic continuation of \eqref{flux1}.
Now, consistent boundary conditions require that this flux is identically zero, in agreement with the dynamical evolution \cite{Ishibashi:2004wx}. Dirichlet boundary conditions $g^{(0)}_{AB} =\mathring q_{AB}$, where $\mathring  q_{AB}$ is the unit metric on the sphere, amount to the standard boundary conditions of asymptotically AdS$_4$ spacetimes \cite{Henneaux:1985tv} and the $\Lambda$-BMS$_4$ group reduces to the standard $SO(3,2)$ group. In this paper, we show that the additional Neumann boundary conditions $J^{AB} = 0$ reduce the $\Lambda$-BMS$_4$ group to the group of temporal shifts $\mathbb R$ times the group of 2-dimensional area-preserving diffeomorphisms $\mathcal A$. Moreover, we show that the associated charges are finite, conserved, integrable and generically non-vanishing. We therefore find new mixed Dirichlet-Neumann boundary conditions with $\mathbb R \times \mathcal A$ as asymptotic symmetry group. This is the third result of this paper. Notably, both Dirichlet and Neumann boundary conditions for the linearized spin 2 field in AdS$_4$ are consistent with a unitary symplectic structure \cite{Ishibashi:2004wx} (see also \cite{Compere:2008us,Andrade:2011dg}). This suggests that our mixed Dirichlet-Neumann boundary conditions will also be unitary though a proof of this conjecture lies beyond the scope of this paper. In the context of holography, Neumann boundary conditions correspond to an electric-magnetic deformation of the dual theory \cite{deHaro:2008gp,Bakas:2008zg}. Mixed Neumann-Dirichlet boundary conditions therefore correspond to a partial electric-magnetic deformation that remains to be understood.

While the Kerr-AdS black hole is not part of the phase space associated to our new boundary conditions, we show that the latter includes analytic stationary solutions with mass and angular momentum. If some of these solutions are regular in the bulk of spacetime, they could bear on the cosmic censorship conjecture in vacuum asymptotically AdS$_4$ spacetimes \cite{Markeviciute:2017jcp,Crisford:2018qkz}.

Let us finally comment on how our results extend the literature. Our construction extends some results of the recent work \cite{Poole:2018koa} in two ways: we keep the area $\sqrt{|g_{(0)}|}$ arbitrary throughout our computation and we do not assume axial and reflection symmetry (which also completes the work of \cite{He:2015wfa}). We therefore provide the most general solution to Einstein's equations with $\Lambda \neq 0$ in Bondi gauge. Second, we extend the dictionary between Bondi gauge and Fefferman-Graham gauge derived in \cite{Poole:2018koa} in a way that preserves covariance with respect to the boundary metric. Our derivation also provides a generalization of the three-dimensional analysis \cite{Barnich:2012aw} to four spacetime dimensions, and a generalization of the four-dimensional asymptotically flat analyses \cite{Bondi:1962px,Sachs:1962wk,Tamburino:1966zz,Winicour1985,Barnich:2009se,Barnich:2010eb,Compere:2018ylh} in the presence of a cosmological constant.



\section{Bondi gauge in (A)dS$_{\mathbf{4}}$}
\label{sec2}

In the following we derive the general solution space of Einstein gravity coupled to a cosmological constant $\Lambda$ of either sign in Bondi gauge. We contrast this solution space with the asymptotically flat case $\Lambda = 0$.

\subsection{Bondi gauge and residual transformations}

The Bondi gauge in $3+1$ dimensions can be defined in the spacetime coordinates $(u, r, x^A)$, where $u$ labels null hypersurfaces, $r$ is the affine parameter along the generating null geodesics and $x^A = (\theta, \phi)$ are transverse angular coordinates. The general ansatz for the metric is given by
\begin{equation}
ds^2 = e^{2\beta} \frac{V}{r} du^2 - 2 e^{2\beta}du dr + g_{AB} (dx^A - U^A du)(dx^B - U^B du)
\label{bondi gauge}
\end{equation} where $\beta$, $U^A$, $g_{AB}$ and $V$ are arbitrary functions of the coordinates. The $2$-dimensional metric $g_{AB}$ satisfies the determinant condition
\begin{equation}
\partial_r \left(\frac{\det (g_{AB})}{r^4} \right) = 0 \Leftrightarrow \det (g_{AB}) = r^4 \chi(u,x^A) \label{eq:DetCond} ,
\end{equation} 
where $\chi(u,x^A)$ is not fixed. Any metric can be written in this gauge. For example, global (A)dS$_4$ is obtained by choosing $\beta = 0$, $U^A = 0$, $V/r = (\Lambda r^2/3)-1$, $g_{AB} = r^2 \mathring q_{AB}$, where $\mathring q_{AB}$ is the unit round-sphere metric. 

Infinitesimal diffeomorphisms preserving the Bondi gauge are generated by vector fields $\xi^\mu$ satisfying
\begin{equation}
\mathcal{L}_\xi g_{rr} = 0, \quad \mathcal{L}_\xi g_{rA} = 0, \quad g^{AB} \mathcal{L}_\xi g_{AB} = 4 \omega(u, x^A).
\label{eq:GaugeConstraints}
\end{equation} The prefactor of $4$ is introduced for convenience. The last condition is a consequence of the determinant condition \eqref{eq:DetCond}. Indeed,
\begin{equation}
g^{AB} \mathcal{L}_\xi g_{AB} = \mathcal{L}_\xi [\ln \det(g_{AB})] \overset{\eqref{eq:DetCond}}{=} \frac{4}{r} \xi^r + \chi^{-1} \mathcal{L}_\xi \chi.
\end{equation} Since $\chi$ is an arbitrary function of $(u,x^A)$, the right-hand side has to be an arbitrary function of $(u,x^A)$, that we write $4\omega(u,x^A)$.  From \eqref{eq:GaugeConstraints}, we deduce
\begin{equation}
\begin{split}
\xi^u &= f, \\
\xi^A &= Y^A + I^A, \quad I^A = -\partial_B f \int_r^\infty dr'  (e^{2 \beta} g^{AB}),\\
\xi^r &= - \frac{r}{2} (\mathcal{D}_A Y^A - 2 \omega + \mathcal{D}_A I^A - \partial_B f U^B + \frac{1}{2} f g^{-1} \partial_u g) ,\\
\end{split}
\label{eq:xir}
\end{equation} 
where $\partial_r f = 0 = \partial_r Y^A$, and $g= \det (g_{AB})$. The covariant derivative $\mathcal{D}_A$ is associated with the $2$-dimensional metric $g_{AB}$. The residual gauge transformations are parametrized by the 4 functions $\omega$, $f$ and $Y^A$ of $(u,x^A)$.

\subsection{Procedure to resolve Einstein's equations}

We solve Einstein's equations $G_{\mu\nu} + \Lambda g_{\mu\nu} = 0$ for pure gravity in Bondi gauge. We follow the integration scheme and the notations of \cite{Barnich:2010eb}. In particular, we use the Christoffel symbols that have been derived in this reference.

\subsubsection{Minimal fall-off requirements}

We impose the fall-off condition $g_{AB} = \mathcal{O}(r^2)$. We assume an analytic expansion for $g_{AB}$, namely 
\begin{equation}
g_{AB} = r^2 \, q_{AB}  + r\, C_{AB} + D_{AB} + \frac{1}{r} \, E_{AB} + \frac{1}{r^2} \, F_{AB} + \mathcal{O}(r^{-3})\label{eq:gABFallOff}
\end{equation} 
where each term involves a symmetric tensor whose components are arbitrary functions of $(u,x^C)$. For $\Lambda \neq 0$, the Fefferman-Graham theorem  \cite{Starobinsky:1982mr,Fefferman:1985aa,Skenderis:2002wp,2007arXiv0710.0919F,Papadimitriou:2010as} together with the map between Fefferman-Graham gauge and Bondi gauge, derived in Appendix \ref{app:chgt}, ensures that the expansion \eqref{eq:gABFallOff} leads to the most general solution to the vacuum Einstein equations. For $\Lambda = 0$, the analytic expansion \eqref{eq:gABFallOff} is an hypothesis since additional logarithmic branches might occur \cite{Winicour1985,Chrusciel:1993hx,ValienteKroon:1998vn}. 

This fall-off condition does not impose any constraint on the generators of residual diffeomorphisms \eqref{eq:xir}. In the following, upper case Latin indices will be lowered and raised by the $2$-dimensional metric $q_{AB}$ and its inverse. The gauge condition \eqref{eq:DetCond} implies $g^{AB}\p_r g_{AB}=4/r$ which imposes successively that $\chi = \det (q_{AB})$, $q^{AB} C_{AB} = 0$ and
\begin{equation}
\begin{split}
&D_{AB} = \frac{1}{4} q_{AB} C^{CD} C_{CD} + \mathcal{D}_{AB} (u,x^C),  \\
&E_{AB} = \frac{1}{2} q_{AB} \mathcal{D}_{CD}C^{CD} + \mathcal{E}_{AB} (u,x^C), \\
&F_{AB} = \frac{1}{2} q_{AB} \Big[ C^{CD}\mathcal{E}_{CD} + \frac{1}{2} \mathcal{D}^{CD}\mathcal{D}_{CD} - \frac{1}{32} (C^{CD}C_{CD})^2 \Big] + \mathcal{F}_{AB}(u,x^C),
\end{split}
\end{equation}
with $q^{AB} \mathcal{D}_{AB} = q^{AB} \mathcal{E}_{AB} = q^{AB} \mathcal{F}_{AB} = 0$. 

\subsubsection{Organization of Einstein's equations}

We organize the equations of motion as follows. First, we solve the equations that do not involve the cosmological constant. The radial constraint $G_{rr} = R_{rr} = 0$ fixes the $r$-dependence of $\beta$, while the cross-term constraint $G_{rA} = R_{rA} = 0$ fixes the $r$-dependence of $U^A$. 

Next, we treat the equations that do depend upon $\Lambda$. The equation $G_{ur} + \Lambda g_{ur} = 0$ determines the $r$-dependence of $V/r$ in terms of the previous variables. Noticing that $R = g^{\mu\nu} R_{\mu\nu} = 2 g^{ur} R_{ur} + g^{rr} R_{rr} + 2 g^{rA} R_{rA} + g^{AB} R_{AB}$, and taking into account that $R_{rr} = 0 = R_{rA}$, one gets $G_{ur} + \Lambda g_{ur} = R_{ur} - \frac{1}{2} g_{ur} R + \Lambda g_{ur} = \frac{1}{2} g_{ur} (2\Lambda- g^{AB} R_{AB} )= 0$ so that we can solve equivalently $g^{AB} R_{AB} = 2\Lambda$. 

Next, we concentrate on the pure angular equation, $G_{AB} + \Lambda g_{AB} = 0$, which can be splitted into a tracefree part 
\begin{equation}
G_{AB} - \frac{1}{2} g_{AB} \, g^{CD}G_{CD} = 0 \label{eq:GABTF}
\end{equation}
and a pure-trace part
\begin{equation}
g^{CD} G_{CD} + 2\Lambda = 0. \label{eq:GABTFull}
\end{equation}
Consider the Bianchi identities $\nabla_\mu G^{\mu\nu} = 0$ which can be rewritten as
\begin{equation}
2 \sqrt{-g} \nabla_\mu G^\mu_\nu = 2 \partial_\mu (\sqrt{-g}G^\mu_\nu) - \sqrt{-g} G^{\mu\lambda} \partial_\nu g_{\mu\lambda} = 0. \label{rewrite Bianchi}
\end{equation}
Since $\partial_\nu g_{\mu\lambda} = - g_{\mu\alpha}g_{\lambda\beta}\partial_\nu g^{\alpha\beta}$, we have
\begin{equation}
2 \partial_\mu (\sqrt{-g}G^\mu_\nu) + \sqrt{-g} G_{\mu\lambda} \partial_\nu g^{\mu\lambda} = 0.
\end{equation}
Taking $\nu = r$ and noting that $G_{r\alpha} +\Lambda g_{r\alpha} = 0$ have already been solved, one gets
\begin{equation}
G_{AB} \partial_r g^{AB} = \frac{4\Lambda}{r}. \label{eq:int1}
\end{equation}
Recalling that \eqref{eq:GABTF} holds, and that the determinant condition implies that $g^{AB}\partial_r g_{AB} = 4/r$, we see that  \eqref{eq:int1} is equivalent to \eqref{eq:GABTFull}. As a consequence, the equation $G_{AB} + \Lambda g_{AB} = 0$ is completely obeyed if \eqref{eq:GABTF} is solved. Indeed, once the tracefree part \eqref{eq:GABTF} has been set to zero, the tracefull part \eqref{eq:GABTFull} is automatically constrained by the Bianchi identity. Another way to see this is as follows. Imposing that $G_{r\alpha} +\Lambda g_{r\alpha} = 0$ holds, \eqref{eq:GABTF} is equivalent to
\begin{equation}
(M^{TF})^A_{\,\,B} \equiv M^A_{\,\,B} - \frac{1}{2} \delta^A_B M^C_{\,\,C} = 0, \quad M^A_{\,\,B} \equiv g^{AC}R_{CB}, \label{eq:MABTF}
\end{equation}
since the trace part of $M^A_{\,\,B}$ has already been set to $2\Lambda$ in order to fix the radial dependence of $V/r$. 

At this stage, Einstein's equations $(r,r)$, $(r,A)$, $(r,u)$ and $(A,B)$ have been solved. It remains to solve the $(u,u)$ and $(u,A)$ components. Doing so we will derive the evolution equations for the Bondi mass and angular momentum aspects, see Section \ref{partII} below. Expressing the $A$ component of the contracted Bianchi identities \eqref{rewrite Bianchi} yields
\begin{equation}
\partial_r \Big[ r^2 \Big( G_{uA} + \Lambda g_{uA} \Big) \Big] = \partial_r \Big[ r^2 \Big( R_{uA} - \Lambda g_{uA} \Big) \Big] = 0. \\
\end{equation}
The equation of motion $r^2 (G_{uA} + \Lambda g_{uA}) = 0$ can therefore be solved for a single value of $r$. We can isolate the only non-trivial equation to be the $1/r^2$ part of $G_{uA} + \Lambda g_{uA} = 0$. This will determine the evolution of the Bondi angular momentum aspect, denoted by $N^{(\Lambda)}_A(u,x^B)$. Assuming that $G_{uA} + \Lambda g_{uA} = 0$ is solved, the last Bianchi identity \eqref{rewrite Bianchi} for $\nu = u$ becomes
\begin{equation}
\partial_r \Big[ r^2 \Big( G_{uu} + \Lambda g_{uu} \Big) \Big] = \partial_r \Big[ r^2 \Big( R_{uu} - \Lambda g_{uu} \Big) \Big] = 0,
\end{equation}
and the reasoning is similar. We will solve the $r$-independent part of $r^2 (R_{uu} - \Lambda g_{uu})$, which will uncover the equation governing the time evolution of the Bondi mass aspect $M^{(\Lambda)}(u,x^A)$.

\subsection{Solution to the algebraic equations}

We define several auxiliary fields as in \cite{Barnich:2010eb}. Starting from \eqref{eq:gABFallOff}, we can build $k_{AB} = \frac{1}{2} \partial_r g_{AB}$, $l_{AB} = \frac{1}{2} \partial_u g_{AB}$, and $n_A = \frac{1}{2} e^{-2\beta}g_{AB}\partial_r U^B$. The determinant condition \eqref{eq:DetCond} allows us to split the tensors $k_{AB}$ and $l_{AB}$ in leading trace-full parts and subleading trace-free parts as
\begin{equation}
\begin{split}
k^A_B &\equiv g^{AC} k_{BC} = \frac{1}{r} \delta^A_B + \frac{1}{r^2} K^A_B, \qquad K^A_A = 0, \\
l^A_B &\equiv g^{AC} l_{BC} = \frac{1}{2} q^{AC}\partial_u q_{BC} + \frac{1}{r} L^A_B, \qquad L^A_A = 0.
\end{split}
\end{equation}
Note that 
\begin{eqnarray}
l = l^A_A = \frac{1}{2} q^{AB}\partial_u q_{AB} = \partial_u \ln \sqrt{q}.
\end{eqnarray} 
Let us start by solving $R_{rr} = 0$ which leads to
\begin{equation}
\partial_r \beta = -\frac{1}{2r} + \frac{r}{4} k^A_B k^B_A = \frac{1}{4r^3} K^A_B K^B_A.
\end{equation}
Expanding $K^A_B$ in powers of $1/r$, we get
\begin{align}
\beta(u,r,x^A) &= \beta_0 (u,x^A) + \frac{1}{r^2} \Big[ -\frac{1}{32} C^{AB} C_{AB} \Big] + \frac{1}{r^3} \Big[ -\frac{1}{12} C^{AB} \mathcal{D}_{AB} \Big] \label{eq:EOM_beta} \\
&\qquad + \frac{1}{r^4}\Big[ - \frac{3}{32} C^{AB}\mathcal{E}_{AB} - \frac{1}{16} \mathcal{D}^{AB}\mathcal{D}_{AB} + \frac{1}{128} (C^{AB}C_{AB})^2 \Big] + \mathcal{O}(r^{-5}). \nonumber
\end{align}
Up to the integration ``constant'' $\beta_0 (u,x^A)$, the condition \eqref{eq:gABFallOff} uniquely determines  $\beta$. In particular, the $1/r$ order is always zero on-shell. This equation also holds for $\Lambda =0$ but standard asymptotic flatness conditions set $\beta_0 = 0$. We shall keep it arbitrary here. 

Next, we develop $R_{rA} = 0$, which gives
\begin{equation}
\partial_r (r^2 n_A) = r^2 \Big( \partial_r - \frac{2}{r} \Big) \partial_A \beta - \mathcal{D}_B K^B_A.
\end{equation}
We now expand the transverse covariant derivative $\mathcal{D}_A$
\begin{equation}
\Gamma^B_{AC}[g_{AB}] = \Gamma^B_{AC}[q_{AB}] + \frac{1}{r} \Big[ \frac{1}{2} (D_A C^B_C + D_C C^B_A - D^B C_{AC}) \Big] + \mathcal{O}(r^{-2}),
\end{equation}
in terms of the transverse covariant derivative $D_A$ defined with respect to the leading transverse metric $q_{AB}$. This implies in particular that
\begin{equation}
\mathcal{D}_B K^B_A = -\frac{1}{2} D^B C_{AB} + \frac{1}{r} \Big[ -D^B \mathcal{D}_{AB} + \frac{1}{8} \partial_A (C_{BC}C^{BC}) \Big] + \mathcal{O}(r^{-2}).
\end{equation}
Using explicitly \eqref{eq:EOM_beta}, we find
\begin{equation}
n_A = -\partial_A \beta_0 + \frac{1}{r}\Big[ \frac{1}{2}D^B C_{AB} \Big] + \frac{1}{r^2} \Big[ \ln r \,  D^B \mathcal{D}_{AB} + N_A \Big] + \mathcal{o}(r^{-2})
\end{equation}
where $ N_A$ is a second integration ``constant'' (\textit{i.e.} $\partial_r  N_A = 0$), which corresponds to the Bondi angular momentum aspect in the asymptotically flat case. After inverting the definition of $n_A$, integrating one time further on $r$ and raising the index $A$, we end up with
\begin{equation}
\begin{split}
U^A = \,\, & U^A_0(u,x^B) +\overset{(1)}{U^A}(u,x^B) \frac{1}{r} + \overset{(2)}{U^A}(u,x^B) \frac{1}{r^2} \\
&+ \overset{(3)}{U^A}(u,x^B) \frac{1}{r^3} + \overset{(\text{L}3)}{U^A}(u,x^B) \frac{\ln r}{r^3} + \mathcal{o}(r^{-3})
\end{split} \label{eq:EOM_UA}
\end{equation}
with
\begin{eqnarray}
\overset{(1)}{U^A}(u,x^B)\hspace{-6pt} &=&\hspace{-6pt} 2 e^{2\beta_0} \partial^A \beta_0 ,\nonumber \\
\overset{(2)}{U^A}(u,x^B)\hspace{-6pt} &=&\hspace{-6pt} - e^{2\beta_0} \Big[ C^{AB} \partial_B \beta_0 + \frac{1}{2} D_B C^{AB} \Big], \nonumber\\
\overset{(3)}{U^A}(u,x^B)\hspace{-6pt} &=& \hspace{-6pt}- \frac{2}{3} e^{2\beta_0} \Big[ N^A - \frac{1}{2} C^{AB} D^C C_{BC} +   (\partial_B \beta_0 + \frac{1}{3} D_B) \mathcal{D}^{AB} - \frac{3}{16} C_{CD}C^{CD} \partial^A \beta_0  \Big], \nonumber\\
\overset{(\text{L}3)}{U^A}(u,x^B) \hspace{-6pt}&=&\hspace{-6pt} -\frac{2}{3}e^{2\beta_0}D_B \mathcal{D}^{AB}, \label{eq:EOM_UA2}
\end{eqnarray}
where $U^A_0(u,x^B)$  is a new integration ``constant''. Again, this equation also holds if $\Lambda$ is absent, but standard asymptotic flatness sets this additional parameter to zero. As known in standard flat case analysis, the presence of $\mathcal{D}_{AB}$ is responsible of logarithmic terms in the expansion of $U^A$. We will shortly derive that for $\Lambda \neq 0$, $\mathcal{D}_{AB}$ vanishes on-shell.

Given that
\begin{equation}
\begin{split}
M^A_{\,\,B} &= e^{-2\beta} \Big[ (\partial_r + \frac{2}{r}) (l^A_B + k^A_B \frac{V}{r} + \frac{1}{2} \mathcal{D}_B U^A + \frac{1}{2} \mathcal{D}^A U_B) \\
&\qquad\qquad + k^A_C \mathcal{D}_B U^C - k^C_B \mathcal{D}_C U^A + (\partial_u + l)k^A_B + \mathcal{D}_C (U^C k^A_B) \Big] \\
&\qquad + R^A_B[g_{CD}] - 2(\mathcal{D}_B \partial^A \beta + \partial^A \beta \partial_B \beta + n^A n_B),
\end{split}
\end{equation}
we extract the $r$-dependence of $V/r$ thanks to $M^A_{\,\,A} = 2\Lambda$, which reads as
\begin{equation}
\begin{split}
\partial_r V = & - 2 r (l + D_A U^A) + \\ &e^{2\beta} r^2 \Big[ D_A D^A \beta + (n^A - \partial^A \beta) (n_A - \partial_A \beta) - D_A n^A - \frac{1}{2} R[g_{AB}] + \Lambda \Big] .
\end{split}
\end{equation}
Taking into account \eqref{eq:gABFallOff}, \eqref{eq:EOM_beta} and \eqref{eq:EOM_UA}, we get after integration on $r$
\begin{align}
\frac{V}{r} = &\frac{\Lambda}{3} e^{2\beta_0} r^2 - r (l + D_A U^A_0) \label{eq:EOMVr} \\
&- e^{2\beta_0} \Big[ \frac{1}{2}\Big( R[q] + \frac{\Lambda}{8}C_{AB} C^{AB} \Big) + 2 D_A \partial^A \beta_0 + 4 \partial_A \beta_0 \partial^A \beta_0 \Big] + \frac{2  M}{r} + \mathcal{o}(r^{-1}) \nonumber 
\end{align}
where $ M(u,x^A)$ is an integration ``constant'' which, in flat asymptotics, is recognized as the Bondi mass aspect. 

Afterwards, we solve \eqref{eq:MABTF} order by order, which provides us the constraints imposed on each independent order of $g_{AB}$. The leading $\mathcal{O}(r^{-1})$ order of that equation yields
\begin{equation}
\frac{\Lambda}{3} C_{AB} = e^{-2\beta_0} \Big[ (\partial_u - l) q_{AB} + 2 D_{(A} U^0_{B)} - D^C U^0_C q_{AB} \Big].
\label{eq:CAB}
\end{equation}
This result shows that there is a discrete bifurcation between the asymptotically flat case and the case $\Lambda \neq 0$. Indeed, when $\Lambda = 0$, the left-hand-side vanishes, which leads to a constraint on the time-dependence of $q_{AB}$. As a consequence, the field $q_{AB}$ is constrained while $C_{AB}$ is completely free and interpreted as the shear. For (A)dS$_4$ asymptotics, $C_{AB}$ is entirely determined by $q_{AB}$ and $U^A_0$, while the boundary metric $q_{AB}=q_{AB}(u,x^A)$ is left completely undetermined by the equations of motion. This is consistent with previous analyses \cite{Ashtekar:2014zfa,He:2015wfa,Saw:2017amv,He:2018ikd,Poole:2018koa}.

Going to $\mathcal{O}(r^{-2})$, we get
\begin{equation}
\frac{\Lambda}{3} \mathcal{D}_{AB} = 0,\label{eq:DAB}
\end{equation}
which removes the logarithmic term in \eqref{eq:EOM_UA} for $\Lambda \neq 0$, but not for $\Lambda = 0$. The condition at next $\mathcal{O}(r^{-3})$ order
\begin{equation}
\partial_u \mathcal{D}_{AB} + U_0^C D_C \mathcal{D}_{AB} + 2 \mathcal{D}_{C(A} D_{B)}U_0^C = 0,
\end{equation}
is thus trivial for $\Lambda \neq 0$, but reduces to $\partial_u \mathcal{D}_{AB} = 0$ in the flat limit, consistently with previous results. 

Using an iterative argument as in \cite{Poole:2018koa}, we now make the following observation. If we decompose $g_{AB} = r^2 \sum_{n\geq 0} g_{AB}^{(n)} r^{-n}$, we see that the iterative solution of \eqref{eq:MABTF} organizes itself as $\Lambda g_{AB}^{(n)} = \partial_u g_{AB}^{(n-1)} + (...)$ at order $\mathcal{O}(r^{-n})$, $n\in\mathbb{N}_0$. Accordingly, the form of $\mathcal{E}_{AB}$ should have been fixed by the equation found at $\mathcal{O}(r^{-3})$, but it is not the case, since both contributions of $\mathcal{E}_{AB}$ cancel between $G_{AB}$ and $\Lambda g_{AB}$. Moreover, the equation $\Lambda g_{AB}^{(4)} = \partial_u g_{AB}^{(3)} + (...)$ at next order turns out to be a constraint for $g_{AB}^{(4)} \sim \mathcal{F}_{AB}$, determined with other subleading data such as $C_{AB}$ or $\partial_u g_{AB}^{(3)} \sim \partial_u \mathcal{E}_{AB}$. It shows that $\mathcal{E}_{AB}$ is a set of two free data on the boundary, built up from two arbitrary functions of $(u,x^A)$. It shows moreover that there is no more data to be uncovered for $\Lambda \neq 0$. This matches with the number of free data of the solution space in Fefferman-Graham gauge, as we will review in Section \ref{sec:FGg}.

As a conclusion, Einstein's equations $(r,r)$, $(r,A)$, $(r,u)$ and $(A,B)$ can be solved iteratively in the asymptotic expansion for $\Lambda \neq 0$. We identified 11 independent functions $\{ \beta_0 (u,x^A)$, $U^A_0 (u,x^B)$, $q_{AB} (u,x^C)$, $ M (u,x^C)$, $ N_A(u,x^C)$, $\mathcal{E}_{AB} (u,x^C)\}$ that determine the asymptotic solution. We will see in Section \ref{partII} that the remaining equations are equivalent to evolution equations for $ M(u,x^A)$ and $ N_A(u,x^B)$. This contrasts with the asymptotically flat case $\Lambda = 0$ where an infinite series of functions appear in the radial expansion, see e.g. \cite{Barnich:2010eb}.

\subsection{Boundary gauge fixing}
\label{sec:bndg}

In this section, we will simplify our analysis by imposing a (codimension one) boundary gauge fixing.  Let us consider the pullback of the most general Bondi metric satisfying \eqref{eq:gABFallOff} to the boundary $\mathscr{I} \equiv \lbrace r \to \infty \rbrace$, 
\begin{equation}
\left. ds^2 \right|_{\mathscr{I}} = \Big[ \frac{\Lambda}{3}e^{4\beta_0} + U_0^A U^0_A \Big] du^2 - 2 U_A^0 du dx^A + q_{AB} dx^A dx^B.
\end{equation}
We will use the boundary gauge freedom to reach the gauge
\begin{equation}
\beta_0 = 0,\quad U^A_0 = 0,\quad \sqrt{q} = \sqrt{\bar q}\label{bndgauge}
\end{equation}
where $\sqrt{\bar q}$ is a fixed area of the 2-dimensional transverse space spanned by $x^A$. This gauge is a temporal boundary gauge for $\Lambda < 0$, a radial boundary gauge for $\Lambda > 0$ and a null boundary gauge for $\Lambda = 0$ with $g_{ur} = -1+ \mathcal{O}(r^{-1})$ in \eqref{bondi gauge}. 

 Intuitively, this amounts to use the gauge freedom at the boundary $\mathscr{I}$, to eliminate three pure-gauge degrees of freedom thanks to a diffeomorphism defined intrinsically on $\mathscr{I}$ and lifted to the bulk in order to preserve Bondi gauge. Such a transformation also involves a Weyl rescaling of the boundary metric, as can be seen from \eqref{eq:xir}, which consists in a redefinition of the luminosity distance $r$ by an arbitrary factor depending on $(u,x^A)$. We can use this Weyl rescaling to gauge-fix one further quantity in the boundary metric, namely the area of the transverse space. Let us now  provide the details. 
 
Computing the Lie derivative on the Bondi metric on-shell and retaining only the leading $\mathcal{O}(r^2)$ terms, we get the transformation laws of the boundary fields $q_{AB}$, $\beta_0$ and $U^A_0$ under the set of residual gauge transformations \eqref{eq:xir}:
\begin{align}
\delta_\xi q_{AB} &= f(\partial_u - l)q_{AB} + (\mathcal{L}_Y - D_C Y^C +2\omega)q_{AB} \nonumber \\
&\quad - 2 (U_{(A}^0 \partial_{B)}f - \frac{1}{2} q_{AB} U^C_0 \partial_C f),\label{eq56} \\
\delta_\xi \beta_0 &= (f \partial_u + \mathcal{L}_Y) \beta_0 + \frac{1}{2}\Big[ \partial_u - \frac{1}{2}l + \frac{3}{2} U^A_0 \partial_A \Big] f - \frac{1}{4} (D_A Y^A  - 2 \omega),\label{eq57} \\
\delta_\xi U^A_0 &= (f\partial_u + \mathcal{L}_Y) U_0^A - \Big[ \partial_u Y^A - \frac{1}{\ell^2} e^{4\beta_0} q^{AB} \partial_B f \Big] + U_0^A (\partial_u f + U_0^B \partial_B f).\label{eq58}
\end{align}
The first equation implies implies that $q^{AB}\delta_\xi q_{AB} = 4\omega$. We can therefore adjust the Weyl generator $\omega$ in order to reach the gauge $ \sqrt{q} = \sqrt{\bar q}$. The form of the infinitesimal transformations \eqref{eq57}-\eqref{eq58} involves $\p_u f$ and $\p_u Y^A$. This ensures that a finite gauge transformation labelled by $f,Y^A$ can be found by integration over $u$ in order to reach $\beta_0=0$, $U^A_0=0$, at least in a local patch. As a result, the conditions \eqref{bndgauge} can be reached by gauge fixing, at least locally. The vanishing of the inhomogeneous contributions in the transformation laws \eqref{eq57}-\eqref{eq58} constrains parameters $f,Y^A$ and reduces the set of allowed vectors among \eqref{eq:xir}. The remaining residual transformations will be studied in Section \ref{sec:LBMS}.

\subsection{Constraint equations as Bondi evolution equations}
\label{partII}

Assuming the gauge fixing conditions \eqref{bndgauge}, we are now ready to present the evolution equations that follow from the remaining Einstein equations. We will moreover suppose that $\mathcal{D}_{AB} = 0$ in the case $\Lambda = 0$ in order to simplify our computation. As justified before, the $\mathcal{O}(r^0)$ part of $r^2 (R_{uA} - \Lambda g_{uA}) = 0$ will fix the temporal evolution of $N_A$. From the Christoffel symbols, we can develop the first term as
\begin{align}
R_{uA} = &-(\partial_u - l) \partial_A \beta - \partial_A l - (\partial_u + l) n_A  \\
&+ n_B \mathcal{D}^B U_A - \partial_B \beta \mathcal{D}_A U^B + 2 U^B (\partial_A \beta \partial_B \beta + n_A n_B) \nonumber \\
&+ \mathcal{D}_B \Big[ l^B_A + \frac{1}{2} ( \mathcal{D}^B U_A - \mathcal{D}_A U^B ) + U^B (\partial_A \beta - n_A) \Big] + 2 n_B l^B_A \nonumber \\
&- \frac{1}{2}(\partial_r + 2\partial_r \beta + \frac{2}{r})  \partial_A \frac{V}{r} - \frac{V}{r} (\partial_r + \frac{2}{r})n_A + k_A^B (\partial_B \frac{V}{r} + 2 \frac{V}{r} n_B) \nonumber \\
&- e^{-2\beta} (\partial_r + \frac{2}{r}) \Big[ U^B(l_{AB} + \frac{V}{r} k_{AB} + \mathcal{D}_{(A}U_{B)}) \Big] \nonumber \\
&- e^{-2\beta} U^B \Big[ (\partial_u + l) k_{AB} - 4 l^C_{(A} k_{B)C} - 2 k^C_A k_{BC} \frac{V}{r} + \mathcal{D}_C (k_{AB}U^C) - 2 k_{C(A}\mathcal{D}^C U_{B)} \Big]. \nonumber
\end{align}
Let us emphasize that the $r$-dependence of the fields is not yet explicit in this expression, so the upper case Latin indices are lowered and raised by the full metric $g_{AB}$ and its inverse. Expanding all the fields in power series of $1/r$ in $R_{uA}$ and $\Lambda g_{uA}$ and selecting the $1/r^2$ terms yields
\begin{equation}
(\partial_u + l)  N_A^{(\Lambda)} - \partial_A  M^{(\Lambda)} - \frac{\Lambda}{2} D^B  J_{AB} = 0. \label{eq:EvolutionNA}
\end{equation}
Here, we defined with hindsight the Bondi mass and angular momentum aspects for $\Lambda \neq 0$ as
\begin{align}
M^{(\Lambda)} &=  M + \frac{1}{16} (\partial_u + l)(C_{CD}C^{CD}), \label{eq:hatM} \\
N^{(\Lambda)}_A &=  N_A - \frac{3}{2\Lambda} D^B (N_{AB} - \frac{1}{2} l C_{AB}) - \frac{3}{4} \partial_A (\frac{1}{\Lambda} R[q] - \frac{3}{8}  C_{CD}C^{CD}), \label{eq:hatNA}
\end{align}
and the traceless symmetric tensor $J_{AB}$ ($q^{AB} J_{AB} = 0$) as
\begin{align}
J_{AB} = &-\mathcal{E}_{AB} -\frac{3}{\Lambda^2} \Big[ \partial_u (N_{AB} - \frac{1}{2} lC_{AB})  -\frac{\Lambda}{2} q_{AB} C^{CD}(N_{CD} - \frac{1}{2} l C_{CD}) \Big] \nonumber \\
&\quad +\frac{3}{\Lambda^2} (D_A D_B l - \frac{1}{2} q_{AB} D_C D^C l) \nonumber \\
&\quad -\frac{1}{\Lambda} (D_{(A}D^C C_{B)C} - \frac{1}{2} q_{AB} D^C D^D C_{CD}) \nonumber \\
&\quad +C_{AB} \Big[ \frac{5}{16} C_{CD}C^{CD} + \frac{1}{2\Lambda}R[q]\Big] . \label{eq:hatJAB}
\end{align}
We used the notation $N_{AB} \equiv \partial_u C_{AB}$. This tensor is symmetric and obeys $q^{AB}N_{AB} = \frac{\Lambda}{3} C^{AB}C_{AB}$. When $\Lambda = 0$, $N_{AB}$ is thus traceless and represents the Bondi news tensor. 

We will justify the definitions of Bondi mass and angular momentum aspects in Section \ref{sec:FGg}. Note that $\partial_u q_{AB}$ has been eliminated using \eqref{eq:CAB}. The transformations of these fields under the residual gauge symmetries $\xi$ preserving the Bondi gauge \eqref{bondi gauge} and the boundary gauge \eqref{bndgauge} are given by
\begin{align}
\delta_\xi M^{(\Lambda)} &= [f\partial_u + \mathcal{L}_Y + \frac{3}{2}(D_A Y^A + f l - 2 \omega)]M^{(\Lambda)} - \frac{\Lambda}{3} N_A^{(\Lambda)} \partial^A f, \label{eq:VarM} \\
\delta_\xi N_A^{(\Lambda)} &= [f\partial_u + \mathcal{L}_Y + D_B Y^B + f l - 2 \omega] N_A^{(\Lambda)} + 3 M^{(\Lambda)} \partial_A f + \frac{\Lambda}{2} J_{AB} \partial^B f, \label{eq:VarNA} \\
\delta_\xi J_{AB} &= [f\partial_u + \mathcal{L}_Y + \frac{1}{2}(D_C Y^C + f l - 2 \omega)] J_{AB} \nonumber \\
&\quad\, - \frac{4}{3} (N_{(A}^{(\Lambda)}\partial_{B)}f - \frac{1}{2} N_C^{(\Lambda)} \partial^C f q_{AB}). \label{eq:VarJAB}
\end{align}

The asymptotically flat limit is not trivial in the equation \eqref{eq:EvolutionNA} due to terms $\sim \Lambda^{-1}$ above which we collect here:
\begin{equation}
\begin{split}
-\frac{3}{2\Lambda} \Big[ &(\partial_u + l) D^B (\partial_u C_{AB} - \frac{1}{2} l C_{AB}) - D^B \partial_u (\partial_u C_{AB} - \frac{1}{2} l C_{AB}) \\
&\quad + \frac{1}{2} (\partial_u + l) \partial_A R[q] + D^B (D_A D_B l - \frac{1}{2} D_C D^C l q_{AB}) \Big]. \label{eq:ProblematicTerms}
\end{split}
\end{equation}
There are two subtle steps here needed in order to massage the evolution equation before taking the limit $\Lambda\to 0$. First, we need to develop the remaining $u$-derivatives acting on covariant derivatives and taking the constraint \eqref{eq:CAB} into account, to highlight $\Lambda$ factors. Next, we can extract the trace of $N_{AB}$, which also contains a residual contribution $\sim \Lambda$. We end up with
\begin{equation}
\begin{split}
\eqref{eq:ProblematicTerms} = \,\, &\frac{1}{2} D_C (N_{AB}^{TF} C^{BC}) + \frac{1}{4} N_{BC}^{TF} D_A C^{BC} - \frac{1}{4} D_A D_B D_C C^{BC} \\
&+ \frac{1}{8} C^B_C C^C_B \partial_A l - \frac{3}{16} l \partial_A (C^B_C C^C_B)
\end{split}
\label{eq:Probl2}
\end{equation}
where $N_{AB}^{TF}$ denotes the tracefree part of $N_{AB}$. The following identities turn out to be useful for the computation:
\begin{equation}
\begin{split}
(\partial_u +l) H^{AB} &= q^{AC}\partial_u H_{CD}q^{BD} - l H^{AB} +\frac{2\Lambda}{3} C^{C(A}H^{B)}_C,   \\
(\partial_u+l) (D^B H_{AB}) &= D^B \partial_u H_{AB} - \frac{1}{2} q^{CD}H_{CD}\partial_A l   \\
& \quad\, - \frac{\Lambda}{3}\Big[ D_C (H_{AB}C^{BC}) + \frac{1}{2} H^{BC}D_A C_{BC}\Big], \\
(\partial_u+l) C^{AB}C_{AB} &= 2 N^{AB}C_{AB} - l C_{AB}C^{AB}, \\
(\partial_u + l) \partial_A R[q] &= - (D^B D_B + \frac{1}{2} R[q]) \partial_A l + \frac{\Lambda}{3} D_A D_B D_C C^{BC}
\end{split}
\end{equation}
where $H_{AB}(u,x^C)$ is any symmetric rank 2 transverse tensor. We note that $N_{AB}^{TF} C^{BC} + C_{AB} N^{BC}_{TF} = \delta_A^C C_{BD}N^{BD}_{TF}$, thanks to which the first term of \eqref{eq:Probl2} can be rewritten as
\begin{equation}
\frac{1}{2} D_C (N_{AB}^{TF} C^{BC}) = \frac{1}{4} D_B (N_{AC}^{TF} C^{BC}-C_{AC} N^{BC}_{TF}) + \frac{1}{4} \partial_A (C_{BD}N^{BD}_{TF}).
\end{equation}
We can now present \eqref{eq:EvolutionNA} in a way that makes terms in $\Lambda$ explicit:
\begin{align}
(\partial_u + l)  N_A &- \partial_A  M - \frac{1}{4} C_{AB} \partial^B R[q] - \frac{1}{16} \partial_A (N_{BC}^{TF} C^{BC}) \label{eq:EvolutionNA_Explicit} \\
&- \frac{1}{32} l \partial_A (C_{BC}C^{BC}) +\frac{1}{4} N_{BC}^{TF} D_A C^{BC} + \frac{1}{4} D_B (C^{BC} N_{AC}^{TF} - N^{BC}_{TF} C_{AC}) \nonumber \\
&+\frac{1}{4} D_B (D^B D^C C_{AC} - D_A D_C C^{BC}) + \frac{\Lambda}{2} D^B(\mathcal{E}_{AB} - \frac{7}{96} C^C_D C^D_C C_{AB}) = 0.\nonumber 
\end{align}
As a result, the asymptotically flat limit can be safely taken and \eqref{eq:EvolutionNA_Explicit} reduces to
\begin{align}
(\partial_u + l)  N_A &- \partial_A  M - \frac{1}{4} C_{AB} \partial^B R[q] - \frac{1}{16} \partial_A (N_{BC}C^{BC}) \nonumber \\
&- \frac{1}{32} l \partial_A (C_{BC}C^{BC}) +\frac{1}{4} N_{BC} D_A C^{BC} + \frac{1}{4} D_B (C^{BC} N_{AC} - N^{BC} C_{AC}) \nonumber \\
&+\frac{1}{4} D_B (D^B D^C C_{AC} - D_A D_C C^{BC}) = 0,
\end{align}
which fully agrees with (4.49) of \cite{Barnich:2010eb} after a change of conventions\footnote{The Bondi news tensor is defined in \cite{Barnich:2010eb} as $N_{AB}^{\text{there}} = \partial_u C_{AB} - l C_{AB}$ while we define $N_{AB}^{\text{here}} = \partial_u C_{AB}$.}. Note that $N_{AB} = N_{AB}^{TF}$ when $\Lambda = 0$.

Let us now derive the temporal evolution of $ M$, encoded in the $r$-independent part of $r^2 (R_{uu} - \Lambda g_{uu})=0$. The first term is worked out to be
\begin{equation}
\begin{split}
R_{uu} = \hspace{0.2cm} &(\partial_u + 2 \partial_u \beta + l) \Gamma^u_{uu} + (\partial_r + 2\partial_r \beta + \frac{2}{r}) \Gamma^r_{uu} + (\mathcal{D}_A + 2\partial_A \beta) \Gamma^A_{uu}  \\
&- 2\partial_u^2 \beta - \partial_u l - (\Gamma^u_{uu})^2 - 2 \Gamma^u_{uA}\Gamma^A_{uu} - (\Gamma^r_{ur})^2 -2\Gamma^r_{uA}\Gamma^A_{ur} - \Gamma^A_{uB}\Gamma^B_{uA}
\end{split}
\end{equation}
where all Christoffel symbols can be found in page 26 of \cite{Barnich:2010eb}. We finally get
\begin{equation}
(\partial_u + \frac{3}{2}l) M^{(\Lambda)} + \frac{\Lambda}{6} D^A  N^{(\Lambda)}_A + \frac{\Lambda^2}{24} C_{AB} J^{AB} = 0. \label{eq:EvolutionM}
\end{equation}
Here, the asymptotically flat limit is straightforward and gives
\begin{equation}
\begin{split}
(\partial_u + \frac{3}{2}l )  M +\frac{1}{8} N_{AB} N^{AB} - \frac{1}{8} l N_{AB} C^{AB} + \frac{1}{32} l^2 C_{AB} C^{AB} - \frac{1}{8} D_A D^A R[q] \\
- \frac{1}{4} D_A D_B N^{AB} + \frac{1}{4} C^{AB} D_A D_B l + \frac{1}{4} \partial_{(A} l D_{B)} C^{AB} + \frac{1}{8} l D_A D_B C^{AB} = 0,
\end{split}\label{duM}
\end{equation}
in agreement with (4.50) of \cite{Barnich:2010eb}. 
As a conclusion, in Bondi gauge \eqref{bondi gauge} with fall-off condition \eqref{eq:gABFallOff} and boundary gauge fixing \eqref{bndgauge}, the general solution to Einstein's equations is entirely determined by the 7 free functions of $(u,x^A)$ for the case $\Lambda \neq 0$: $q_{AB}$ with fixed area $\sqrt{\bar q}$, $ M$, $N_A$ and tracefree $J_{AB}$ where $M$ and $N_A$ are constrained by the evolution equations \eqref{eq:EvolutionM} and \eqref{eq:EvolutionNA}. This contrasts with the asymptotically flat case $\Lambda = 0$ where an infinite series of functions
appearing in the radial expansion of $g_{AB}$ have to be specified in order to parametrize the solution, see e.g. \cite{Barnich:2010eb}.

\section{Bondi news and Bondi mass in (A)dS$_4$}

In asymptotically flat spacetimes with a fixed unit boundary metric on the sphere $q_{AB} = \mathring q_{AB}$, the symplectic flux $\omega^r$ at future null infinity $\mathscr I^+$ is given by 
\begin{eqnarray}
\omega^r = \frac{1}{32 \pi G} \int_{\mathscr I^+} du d^2\Omega \, \sqrt{\bar q}\,  \delta N^{AB} \wedge \delta C_{AB}\label{flux}
\end{eqnarray}
where $N_{AB} = \p_u C_{AB}$ is the Bondi news and $C_{AB}$ is the shear. For the equivalent expression in more general asymptotically flat spacetimes, see \cite{Compere:2018ylh}. This symplectic flux identifies the time dependence of $C_{AB}$ as the cause of energy flux leaking through future null infinity. In order to understand the analogue of Bondi news at the future of de Sitter spacetime, we will derive the analogue of the symplectic flux \eqref{flux} when $\Lambda \neq 0$. We will proceed by mapping quantities in Bondi gauge to Fefferman-Graham gauge where the algebra is easier, and the interpretation clearer from a holographic perspective. It will also lead to the identification of the Bondi mass and angular momentum aspects. Our construction extends earlier results obtained in \cite{Poole:2018koa}. We assume $\Lambda \neq 0$ through this entire section.

\subsection{Gravity in Fefferman-Graham gauge}

Fefferman-Graham coordinates are Gaussian normal coordinates centered at the boundary of asymptotically locally (A)dS$_4$ spacetimes \cite{Starobinsky:1982mr,Fefferman:1985aa,Skenderis:2002wp,2007arXiv0710.0919F,Papadimitriou:2010as}. We denote as $\rho$ the expansion coordinate (with dimension inverse length) and $x^a = (t, x^A)$ the other coordinates. We set the boundary $\mathscr I$ (spatial infinity for $\Lambda < 0$ and either future or past timelike infinity for $\Lambda >0$)  at $\rho =0$. For $\Lambda < 0$, $\rho$ is spacelike while for $\Lambda > 0$, $\rho$ is timelike. The metric is given by
\begin{equation}
ds^2 =- \frac{3}{\Lambda}\frac{d\rho^2}{\rho^2} + \gamma_{ab}(\rho,x^c) dx^a dx^b. \label{FG gauge}
\end{equation} 

The infinitesimal diffeomorphisms preserving the Fefferman-Graham gauge are generated by vector fields $\xi^\mu$ satisfying
$\mathcal{L}_\xi g_{\rho \rho} = 0$, $\mathcal{L}_\xi g_{\rho a} = 0$. The first condition leads to the equation $\partial_\rho \xi^\rho = \frac{1}{\rho} \xi^\rho$, which can be solved for $\xi^\rho$ as 
\begin{eqnarray}
\xi^\rho = \sigma (x^a) \rho. \label{AKV 1}
\end{eqnarray}
The second condition leads to the equation $\rho^2 \gamma_{ab} \partial_\rho \xi^b - \frac{3}{\Lambda}\partial_a \xi^\rho =0$, which can be solved for $\xi^a$ as 
\begin{equation}
\xi^a = \xi_0^a (x^b) + \frac{3}{\Lambda}\partial_b \sigma \int_0^\rho \frac{d\rho'}{\rho'} \gamma^{ab}(\rho', x^c).
\label{AKV 2}
\end{equation} 
Assuming $\gamma_{ab}= \mathcal{O} (\rho^{-2})$, the general asymptotic expansion that solves Einstein's equations is analytic,
\begin{equation}
\gamma_{ab} = \frac{1}{\rho^2} \, g_{ab}^{(0)} + \frac{1}{\rho} \, g_{ab}^{(1)} + g_{ab}^{(2)} + \rho \, g_{ab}^{(3)} + \mathcal{O}(\rho^2)
\label{preliminary FG}
\end{equation} where $g_{ab}^{(i)}$ are arbitrary functions of $x^a = (t, x^A)$. We take the convention that $t$ has the dimension of length and $x^A$ are dimensionless. Note that $t$ is spacelike for $\Lambda > 0$. Following the standard holographic dictionary (see e.g. \cite{deHaro:2000vlm}), we call $g_{ab}^{(0)}$ the boundary metric and 
\begin{equation}
T_{ab} = \frac{\sqrt{3|\Lambda|}}{16\pi G} g_{ab}^{(3)}
\end{equation} 
the energy-momentum tensor. Einstein's equations fix $g^{(1)}_{ab} = 0$ and $g^{(2)}_{ab}$ in terms of $g_{ab}^{(0)}$ while all subleading terms in \eqref{preliminary FG} are determined in terms of the free data $g^{(0)}_{ab}$ and $T_{ab}$ satisfying 
\begin{equation}
D_a^{(0)} T^{ab} = 0 , \quad g^{(0)}_{ab} T^{ab} =0.
\label{condition on energy momentum}
\end{equation} 
Here $D^{(0)}_a$ is the covariant derivative with respect to $g_{ab}^{(0)}$ and indices are raised with the inverse metric $g^{ab}_{(0)}$. The variation of the free data under the residual gauge transformations is given by (see also \cite{Papadimitriou:2010as})
\begin{eqnarray}
\delta_\xi g_{ab}^{(0)} &=& \mathcal{L}_{\xi^c_0} g_{ab}^{(0)} - 2 \sigma\, g_{ab}^{(0)},\\
\delta_\xi T_{ab} &=& \mathcal{L}_{\xi^c_0} T_{ab}+ \sigma\, T_{ab}. \label{eq:TransformationTab}
\end{eqnarray}

\subsection{Dictionary between Fefferman-Graham and Bondi gauges}
\label{sec:FGg}

In Appendix \ref{app:chgt}, we establish a coordinate transformation between Fefferman-Graham and Bondi gauges, which extends the procedure used in \cite{Poole:2018koa} to a generic boundary metric. The boundary metric in Fefferman-Graham gauge is related to the functions in Bondi gauge through
\begin{equation}
g_{tt}^{(0)} = \frac{\Lambda}{3} e^{4 \beta_0} + U_0^C U_{C}^0 , \qquad g_{tA}^{(0)} = - U_A^0, \qquad g_{AB}^{(0)} =   q_{AB},
\label{g0 in term of Bondi}
\end{equation} 
where all functions on the right-hand sides are now evaluated as functions of $(t,x^A)$. 

The parameters $\lbrace \sigma,\xi^t_0, \xi^A_0 \rbrace$ of the residual gauge diffeomorphisms in the Fefferman-Graham gauge \eqref{AKV 1} and \eqref{AKV 2} can be related to those of the Bondi gauge appearing in \eqref{eq:xir} through
\begin{equation}
\begin{split}
\xi^t_0 &= f ,\\
\xi^A_0 &= Y^A, \\
\sigma &= \frac{1}{2} (D_A Y^A + f l - U_0^A \partial_A f - 2\omega),
\end{split}
\label{translation parameters}
\end{equation} 
where all functions on the right-hand sides are also evaluated as functions of $(t,x^A)$. 

The boundary gauge fixing \eqref{bndgauge} described in Section \ref{sec:bndg} can now be understood as a gauge fixation of the boundary metric to
\begin{eqnarray}
g^{(0)}_{tt} = \frac{\Lambda}{3},\qquad g^{(0)}_{tA} =0,\qquad \text{det}(g_{(0)})=  \frac{\Lambda}{3} \bar q.\label{bndgauge2}
\end{eqnarray}
For $\Lambda < 0$ (resp. $\Lambda > 0$), this is exactly the temporal (resp. radial) gauge for the boundary metric, with a fixed area form for the 2-dimensional transverse space. 

Let us develop the constraint equations \eqref{condition on energy momentum} after boundary gauge fixing. First, the tracelessness condition determines the trace of $T_{AB}$ to be 
\begin{equation}
q^{AB}T_{AB} = -\frac{3}{\Lambda} T_{tt}.
\end{equation}
We define $T^{TF}_{AB}$ as the tracefree part of $T_{AB}$, i.e. $T_{AB} = T^{TF}_{AB} -\frac{3}{2\Lambda} T_{tt} q_{AB}$. The conservation equation $D_a^{(0)} T^{ab} = 0$ reads as
\begin{equation}
\begin{split}
(\partial_t + \frac{3}{2} l) T_{tt} + \frac{\Lambda}{3} D^A T_{tA} - \frac{\Lambda}{6} \partial_t q_{AB} T^{AB}_{TF} &= 0, \\
(\partial_t + l) T_{tA} - \frac{1}{2} \partial_A T_{tt} + \frac{\Lambda}{3} D^B T^{TF}_{AB} &= 0.
\end{split}
\label{EOM FG}
\end{equation}
Pursuing the change of coordinates to Fefferman-Graham gauge up to fourth order in $\rho$, it can be shown that the stress tensor is given, in terms of Bondi variables, by
\begin{equation}
T_{ab} = \frac{\sqrt{3 |\Lambda|}}{16\pi G} \left[
\begin{array}{cc}
-\frac{4}{3} M^{(\Lambda)} & -\frac{2}{3}  N^{(\Lambda)}_B \\ 
-\frac{2}{3}  N^{(\Lambda)}_A &  J_{AB} + \frac{2}{\Lambda} M^{(\Lambda)} q_{AB}
\end{array} 
\right], \label{eq:RefiningTab}
\end{equation}
where $M^{(\Lambda)} (t,x^A)$ and $N^{(\Lambda)}_A (t,x^B)$ are the boundary fields defined as \eqref{eq:hatM}-\eqref{eq:hatNA} and $J_{AB}$ is precisely the tensor \eqref{eq:hatJAB}, all evaluated as functions of $t$ instead of $u$.  The conservation equations \eqref{EOM FG} are in fact equivalent to \eqref{eq:EvolutionM} and \eqref{eq:EvolutionNA} after using the dictionary \eqref{eq:RefiningTab} and solving $\partial_t q_{AB}$ in terms of $C_{AB}$ using \eqref{eq:CAB}. Morever, we checked that the transformation laws \eqref{eq:VarM}-\eqref{eq:VarJAB} are equivalent to \eqref{eq:TransformationTab}. We therefore identified the Bondi mass aspect $M^{(\Lambda)}$ and the Bondi angular momentum aspect $N^{(\Lambda)}_A$ as the components $T_{tt}$ and $T_{tA}$ of the holographic stress-tensor, up to a normalization constant.

We recall that for standard asymptotically flat spacetimes with a $u$-independent boundary area form $\sqrt{\bar q}$, the Bondi mass is given by $ \int d^2 \Omega \, \sqrt{\bar q} \, M$ and it obeys the Bondi mass loss formula
\begin{equation}
\p_u \int d^2 \Omega \, \sqrt{\bar q} \, M = -\frac{1}{8} \int d^2 \Omega \, \sqrt{\bar q} \, N_{AB} N^{AB} \leq 0 ,
\end{equation}
consistent with \eqref{duM} after using $l=\partial_u \ln \sqrt{q} = 0$. The integrations are performed on a transverse 2-surface at $\mathscr I$. The definition of Bondi mass for $\Lambda \neq 0$ is instead $ \int d^2 \Omega \, \sqrt{\bar q} \, M^{(\Lambda)}$ and it evolves as
\begin{equation}
\p_u  \int d^2 \Omega \, \sqrt{\bar q} \, M^{(\Lambda)}= - \frac{\Lambda^2}{24} \int d^2 \Omega \, \sqrt{\bar q} \, C_{AB} J^{AB} = - \frac{\Lambda}{8} \int d^2 \Omega \, \sqrt{\bar q} \, \p_u q_{AB} J^{AB},
\end{equation}
after using \eqref{eq:CAB} and \eqref{bndgauge}. The right-hand side is not manifestly non-positive. It remains an important problem to investigate whether the Bondi mass decreases with $u$.

\subsection{Symplectic flux at the boundary of (A)dS$_4$}
\label{sec:flux}

The renormalized action for General Relativity in asymptotically locally (A)dS$_4$ spacetimes is given by
\begin{eqnarray}
S = \frac{1}{16 \pi G} \int_{\mathscr M} d^4 x\sqrt{-g} \, (R[g] - 2 \Lambda) + \frac{1}{16 \pi G} \int_{\mathscr I} d^3 x \sqrt{|\gamma|} \, (2 K + \frac{4}{\ell} - \ell R[\gamma]). \label{Sc}
\end{eqnarray} 
where $\ell = \sqrt{|\Lambda|/3}$. Here, $g_{\mu\nu}$ denotes the bulk metric on the spacetime $\mathscr M$, $\gamma_{ab}$ the induced metric at the boundary $\mathscr I$ defined in \eqref{FG gauge} and $K$ the trace of its intrinsic curvature. The bulk term is the normalized Einstein-Hilbert action. The second term is a boundary counterterm that is required in order to have a well-defined variational principle \cite{Balasubramanian:1999re,deHaro:2000vlm,Papadimitriou:2005ii}. 

The variational principle uniquely determines the symplectic structure. The symplectic form is given by \cite{Compere:2008us}
\begin{eqnarray}\label{omegac}
\omega= \omega_{EH}[\delta g, \delta g ; g] - d \omega_{EH}[\delta \gamma , \delta \gamma ; \gamma]
\end{eqnarray}
where $\omega_{EH}$ is the Lee-Wald symplectic structure of the normalized Einstein-Hilbert action \cite{Lee:1990nz,Iyer:1994ys} for, respectively, the bulk metric $g_{\mu\nu}$ and the boundary metric $\gamma_{ab}$. The symplectic form \eqref{omegac} is finite thanks to the counterterm subtraction. The symplectic flux is defined as the finite $\rho$ component of $\omega$, 
\begin{eqnarray}
\omega^\rho = \frac{1}{2\ell^2} \int_{\mathscr I} d^3 x \, \delta \Big(\sqrt{|g_{(0)}|}  T^{ab}\Big) \wedge \delta g^{(0)}_{ab}. 
\label{Simplectic flux}
\end{eqnarray} 
After imposing the boundary gauge fixing \eqref{bndgauge2}, this expression reduces to
\begin{eqnarray}
\omega^\rho = \frac{3}{32 \pi G\ell^4}\int_{\mathscr I} d^3 x\, \sqrt{\bar q}\, \delta J^{AB} \wedge \delta q_{AB}. 
\label{Symplectic flux after bgf}
\end{eqnarray} 
This shows that the information on gravitational energy fluxes passing through $\mathscr{I}$ is entirely contained in the couple $(J^{AB},q_{AB})$. We will therefore call this couple of functions the news for asymptotically locally (A)dS$_4$ spacetimes. The expression \eqref{Symplectic flux after bgf} is the analogue in asymptotically locally (A)dS$_4$ spacetimes of the asymptotically flat expression \eqref{flux}. The role of  Bondi news and shear $(N^{AB},C_{AB})$ is now played by $(J^{AB},q_{AB})$. Note that while $N_{AB} = \partial_u C_{AB}$, there is no such relationship between $J^{AB}$ and $q_{AB}$ in asymptotically locally dS$_4$ spacetimes since otherwise the Cauchy data would be constrained. In asymptotically locally AdS$_4$ spacetimes, the existence of a symplectic flux \eqref{Symplectic flux after bgf} at spatial infinity shows that the dynamics needs to be completed with a boundary condition on $(J^{AB},q_{AB})$. We will discuss the choice of this  boundary condition in Section \ref{sec:newBC}.

\section{The $\Lambda$-BMS$_4$ algebra} 
\label{sec:LBMS}

After considering either Bondi or Fefferman-Graham gauge, and imposing the boundary gauge fixing conditions, respectively \eqref{bndgauge} or \eqref{bndgauge2}, the residual gauge transformations reduce to a set of transformations labelled by three independent functions of the 2-dimensional coordinates $x^A$, as we will now show. 

The derivation can be performed in either Bondi or Fefferman-Graham gauge. We write it in Bondi gauge. We start from \eqref{eq:xir} and impose \eqref{bndgauge}. Using \eqref{eq56}-\eqref{eq57}-\eqref{eq58}, the condition $\delta_\xi \sqrt{q} = 0$ leads to
\begin{equation}
\omega = 0.\label{eq0a}
\end{equation} The condition $\delta_\xi \beta_0 = 0$ leads to 
\begin{equation}
\Big(\partial_u - \frac{1}{2}l\Big)f = \frac{1}{2} D_A Y^A,\label{eq1a}
\end{equation} while $\delta_\xi U_0^A = 0$ gives
\begin{equation}
\partial_u Y^A = -\frac{\Lambda}{3} \partial^A f.\label{eq2a}
\end{equation} 
The solution to \eqref{eq1a}-\eqref{eq2a} admits three integration ``constants'' $T(x^A),V^A(x^B)$, though these cannot be solved explicitly for an arbitrary transverse metric $q_{AB}$ in terms of these functions. In the Fefferman-Graham notation, the equations \eqref{eq0a}-\eqref{eq1a}-\eqref{eq2a} are equivalent to 
\begin{align}
\sigma &= \frac{1}{2}(D_A^{(0)} \xi^A_0 +f l) , \label{sig}\\
\Big(\partial_t - \frac{1}{2} l\Big) \xi^t_0 &= \frac{1}{2} D_A^{(0)}  \xi^A_0, \quad \partial_t \xi^A_0 = -\frac{\Lambda}{3}g^{AB}_{(0)}\partial_B \xi^t_{(0)}.
\end{align}

The asymptotic Killing vectors induce an action on $\mathscr{I}$ generated by the pullback of $\xi$ at the boundary, $\bar{\xi} = f \partial_u + Y^A \partial_A$. These generators satisfy the algebra
\begin{equation}
[\bar{\xi}_1 , \bar{\xi}_2] = \hat{\bar{\xi}},
\end{equation} where $\hat{\bar{\xi}} = \hat{f} \partial_u + \hat{Y}^A \partial_A$ with
\begin{align}
\hat{f} &= Y_1^A \partial_A f_2 + \frac{1}{2} f_1 D_A Y_2^A - (1 \leftrightarrow 2) \label{bms like algebra 1}, \\
\hat{Y}^A &= Y^B_1 \partial_B Y_2^A - \frac{\Lambda}{3} f_1 q^{AB} \partial_B f_2 - (1 \leftrightarrow 2).
\label{bms like algebra 2}
\end{align} 
We call the vectors generated by $T(x^A)$ and $V^A(x^B) $ the supertranslation and superrotation generators, respectively. In the asymptotically flat limit $\Lambda = 0$ and for time-independent transverse metric $q_{AB} = q_{AB}(x^C)$, the functions $Y^A$, $f$ reduce to $Y^A = V^A(x^B)$, $f = T(x^A) + \frac{u}{2} D_A V^A$ and the structure constants reduce to the ones of the extended BMS$_4$ algebra \cite{Campiglia:2014yka}. For $\Lambda \neq 0$, supertranslations do not commute and the structure constants depend explicitly on $q_{AB}$. We therefore find the structure of a Lie algebroid \cite{2000math.....12106F,Lyakhovich:2004kr,Barnich:2010eb,Barnich:2010xq}. We call it the $\Lambda-$BMS$_4$ algebra.

When the transverse metric $q_{AB}$ is equal to the unit round sphere metric $\mathring{q}_{AB}$, the $\Lambda-$BMS$_4$ algebra contains the $SO(3,2)$ algebra for $\Lambda <0$ and the $SO(1,4)$ algebra for $\Lambda > 0$ (see \cite{Barnich:2013sxa}). For the interested reader, we review the explicit expressions of the $SO(3,2)$ Killing vectors of AdS$_4$ in Appendix \ref{app:ExactVectors}.

\section{New boundary conditions for AdS$_4$}
\label{sec:newBC}

The equations of motion for general locally asymptotically AdS$_4$ spacetimes were solved in Bondi gauge in Section \ref{sec2} and related to Fefferman-Graham gauge in Section \ref{sec:FGg}. We found in Section \ref{sec:LBMS} that the residual gauge transformations both in Bondi and Fefferman-Graham gauges consist of the $\Lambda$-BMS$_4$ group after imposing Dirichlet boundary gauge conditions. Now, the set of solutions and residual gauge transformations do not yet form a consistent phase space for asymptotically AdS$_4$ spacetimes without further boundary conditions because of the presence of symplectic flux \eqref{Symplectic flux after bgf}. In this section, we derive new boundary conditions for asymptotically locally AdS$_4$ spacetimes that admit the Schwarzschild-AdS black hole and stationary rotating solutions distinct from the Kerr-AdS black hole. The asymptotic symmetry group is shown to be a subgroup of the  $\Lambda$-BMS$_4$ group consisting of time translations and area-preserving diffeomorphisms. 

\subsection{Definition}

In the case of locally asymptotically AdS spacetimes, a boundary condition is required at the spatial boundary $\mathscr{I}^0_{\text{AdS}}$ as part of the definition of the dynamics \cite{Ishibashi:2004wx}. Such a boundary condition amounts to require that the symplectic flux \eqref{Simplectic flux} at spatial infinity is identically zero.

In the literature, both Dirichlet and Neumann boundary conditions have been studied. On the one hand, Dirichlet boundary conditions  \cite{Henneaux:1985tv} amount to freeze the components of the boundary metric $g^{(0)}_{ab}$ to the ones of the unit cylinder while leaving the holographic stress-tensor $T^{ab}$ free. The resulting asymptotic symmetry group is the group of exact symmetries of AdS$_4$, namely $SO(3,2)$. On the other hand, Neumann boundary conditions \cite{Compere:2008us} freeze the components of $T^{ab}$ while leaving the boundary metric $g^{(0)}_{ab}$ free. The resulting asymptotic symmetry group is empty: all residual gauge transformations have vanishing charges. 

We now present new mixed Dirichlet-Neumann boundary conditions. We first impose the boundary gauge fixing \eqref{bndgauge2}. This is a Dirichlet boundary condition on part of the boundary metric, which is reachable locally by a choice of gauge\footnote{Note that a diversity of boundary conditions reachable by a choice of gauge exist in asymptotically AdS$_3$ spacetimes, see \cite{Brown:1986nw,Compere:2013bya,Troessaert:2013fma,Avery:2013dja,Grumiller:2016pqb,Perez:2016vqo,Grumiller:2017sjh}.}. The symplectic flux at the spatial boundary is then given by \eqref{Symplectic flux after bgf}. We now further impose the Neumann boundary conditions 
\begin{equation}
J^{AB} = 0.  \label{BCads}
\end{equation}
This cancels the symplectic flux, as required. The boundary condition \eqref{BCads} restricts the phase space of solutions. For definiteness, we will further choose the area form of the transverse space $\bar q$ to be $t$-independent, which implies $l = \p_u \ln \sqrt{\bar q} =0$.

\subsection{Asymptotic symmetry algebra}

Let us now derive the asymptotic symmetries preserving the boundary conditions and derive the associated charge algebra. 

The boundary gauge fixing \eqref{bndgauge2} is preserved by the $\Lambda-$BMS$_4$ group of residual gauge transformations as derived in Section  \ref{sec:LBMS}. We will now show that the boundary condition \ref{BCads} further reduces the $\Lambda-$BMS$_4$ group to the direct product $\mathbb{R} \times \mathcal{A}$ where $\mathbb{R}$ are the time translations and $\mathcal{A}$ is the group of 2-dimensional area-preserving diffeomorphisms. We will show that the charges associated to this asymptotic symmetry group are finite, integrable, conserved and generically non-vanishing on the phase space.  

The variation of $J_{AB}$ is given by 
\begin{equation}
\begin{split}
\delta_\xi {J}_{AB} &= (\xi_0^t \partial_t + \mathcal{L}_{\xi_0^C} + \sigma) {J}_{AB} - \frac{4}{3} \Big[  N_{(A} \partial_{B)}\xi_0^t - \frac{1}{2} N_{C} \partial^C \xi_0^t q_{AB} \Big] \\
&\overset{\eqref{translation parameters}}{=} \Big[ \xi_0^t \partial_t + \mathcal{L}_{\xi_0^C} + \frac{1}{2} D_A \xi^A_0 \Big] {J}_{AB} - \frac{4}{3} \Big[  N_{(A} \partial_{B)}\xi_0^t - \frac{1}{2} N_{C} \partial^C \xi_0^t q_{AB} \Big].
\end{split}
\end{equation} 
We recall that  $D_A$ is the covariant derivative with respect to the transverse metric $g^{(0)}_{AB} = q_{AB}$. Imposing $\delta_\xi J_{AB} = 0$ leads to the following constraint on the residual gauge diffeomorphisms given in equation \eqref{AKV 2}:
\begin{equation}
\partial_A \xi_0^t =0.
\label{angle independance}
\end{equation} Therefore, the asymptotic symmetry generators satisfy the relations
\begin{equation}
\partial_t \xi^t_0 = \frac{1}{2} D_A \xi^A_{0} , \qquad \partial_t \xi^A_0 = 0.
\end{equation} The second equation implies $\xi^A_0 = V^A(x^B)$, while the first gives
\begin{equation}
\xi^t_0 = T + \frac{t}{2} D_A V^A
\end{equation} where $T$ is a constant by virtue of \eqref{angle independance}, and $D_A V^A \equiv c$ where $c$ is also a constant. Using Helmholtz's theorem, the vector $V^A$ can be decomposed into a divergence-free and a curl-free part as $V^A = \epsilon^{AB} \p_B \Phi + q^{AB} \p_B \Psi$ where $\Psi$ and $\Phi$ are functions of $x^C$. Injecting this expression for $V^A$ into this equation gives $D_A D^A \Psi = c$. This equation admits a solution if and only if $c=0$ which is given by $\Psi = 0$. Therefore, the asymptotic symmetry generators are given by 
\begin{equation}
\xi^t_0 = T, \qquad \xi^A_0 =  \epsilon^{AB} \p_B \Phi(x^C)
\end{equation} where $T$ is a constant and $\Phi(x^C)$ is arbitrary. Writing $\bar{\xi} = \bar \xi^a \p_a =  T\partial_t + \epsilon^{AB} \p_B \Phi \partial_A$, we have $[\bar{\xi}_1,  \bar{\xi}_2 ] = \hat{\bar{\xi}}$ where
\begin{equation}
\hat{T} = 0, \qquad \hat{\Phi} = \epsilon^{AB}\p_A \Phi_2 \p_B \Phi_1. 
\label{algebra are preserving}
\end{equation} Hence, after imposing the boundary condition \eqref{BCads}, the $\Lambda-$BMS$_4$ algebra reduces to the $\mathbb{R} \oplus \mathcal{A}$ algebra where $\mathbb{R}$ denotes the abelian time translations and $\mathcal{A}$ is the algebra of 2-dimensional area-preserving diffeomorphisms. The latter symmetries are an infinite-dimensional extension of the $SO(3)$ rotations. 

Let us now obtain the associated charges. The fluctuations of the boundary metric components $q_{AB}$ require a renormalization of the bulk Einstein action which is given by \eqref{Sc}. Indeed, the variation of the renormalized action is 
\begin{eqnarray}
\delta S =\frac{1}{2} \int_{\mathscr I} d^3 x \, \sqrt{-g_{(0)}} \,  T^{ab} \delta g^{(0)}_{ab} = \frac{3}{32 \pi G\ell^2 }\int_{\mathscr I} d^3 x \, \sqrt{|\bar q|} \,  J^{AB}  \delta q_{AB}=0 . 
\end{eqnarray}
The boundary Einstein action (with opposite sign with respect to the bulk action) gives a contribution to the symplectic structure given in \eqref{omegac}. It also gives a contribution to the infinitesimal surface charges, as studied in \cite{Compere:2008us}. The infinitesimal generators have a vanishing Weyl transformation, $\sigma = 0$ \eqref{sig}, as a consequence of $l=0$ and $D_A \xi_0^A = 0$. For any residual gauge transformation which consists of an arbitrary boundary diffeomorphism $\bar \xi^a$ and vanishing Weyl transformation, the variation of the renormalized charges are finite and given by \cite{Compere:2008us} 
\begin{equation}
\oint_{S^\infty}\ndelta Q_\xi [g, \delta g] =  \delta \oint_{S^\infty}d^2 \Omega \,(\sqrt{-g_{(0)}} \, T_{ab} n^a[g_{(0)}] \bar \xi^b  ) - \oint_{S^\infty} i_{\bar \xi} \Theta^{(0)}\label{exprc}
\end{equation} where $S_\infty$ is a $2$-surface which is a section of the boundary space-time, $n^a$ is the unit normal vector of $S_\infty$ at the boundary and $\Theta^{(0)}= \frac{1}{2}\sqrt{-g_{(0)}} \, T^{ab} \delta g^{(0)}_{ab} d^3x$. After imposing the boundary gauge fixing \eqref{bndgauge2} and the boundary condition \eqref{BCads}, the last term in \eqref{exprc} vanishes. As a result, the charges are integrable. The integrated charges reduce to 
\begin{equation}
\oint_{S^\infty} Q_{\xi(T,\Phi)} [g] = \oint_{S^\infty} d^2 \Omega \, \sqrt{\bar q} \, [ T^t_{\; \, \, t}\, T + T^t_{\; \, \, A} \epsilon^{AB} \p_B \Phi ] .
\label{charge expression}
\end{equation} From this expression, we see that the charges associated to the symmetry $\mathbb{R} \oplus \mathcal{A}$ are generically non-vanishing. Taking $T = 1$ and $\Phi = 0$ gives the energy. The first harmonic modes of $\Phi$ give the angular momenta, while the higher modes give an infinite tower of charges. Using the \eqref{algebra are preserving} and \eqref{EOM FG}, a simple computation shows that the charges \eqref{charge expression} satisfy the algebra
\begin{equation}
-\delta_{\xi (T_1, \Phi_1)} Q_{\xi (T_2, \Phi_2)} = Q_{\xi (\hat T,\hat \Phi)} .
\end{equation} The charges form a representation of $\mathbb{R} \oplus \mathcal{A}$ without central extension.

\subsection{Stationary solutions}

Let us now study the stationary sector of the phase space associated with the boundary conditions. The AdS$_4$-Schwarzschild solution is included in the phase space. Indeed, AdS$_4$-Schwarzschild can be set in Fefferman-Graham gauge, which allows to identify $q_{AB} = \mathring q_{AB}$ the unit metric on the sphere, as well as $T^t_{\; \, \, t} = \frac{M}{4 \pi G}$, $T_{tA} = 0$ and $T_{AB} = 0$, which finally implies $J_{AB} = 0$. 

The boundary metric and holographic stress-tensor of Kerr-AdS$_4$ are given in the conformally flat frame by \cite{Awad:1999xx,Bhattacharyya:2007vs,Bhattacharyya:2008ji}
\begin{eqnarray}
g^{(0)}_{ab}dx^a dx^b &=& -\ell^{-2}dt^2 + d\theta^2 +  \sin^2\theta d\phi^2,\label{flat} \\
T^{ab} &=& T_{\text{Kerr}}^{ab}  \equiv - \frac{m \gamma^3\ell }{8 \pi} (3 u^a u^b + g_{(0)}^{ab}),
\end{eqnarray}
where $\Xi = 1-a^2 \ell^{-2}$ and 
\begin{eqnarray}
u^a \p_a = \gamma \ell (\p_t +\frac{a}{\ell^2} \p_\phi),\qquad \gamma^{-1} \equiv \sqrt{1-\frac{a^2}{\ell^2} \sin^2\theta}. 
\end{eqnarray} 
The mass and angular momentum are $M = \int \sqrt{\bar q} \, T^t_{\,\, t} = \frac{m}{\Xi^2}$,  $J = Ma = -\int \sqrt{\bar q} \, T^t_{\,\,\phi} = \frac{ma}{\Xi^2}$. We observe that $J_{AB} \neq 0$. Therefore, the Kerr-AdS$_4$ solution is not included in the phase space. However, it is possible to obtain a stationary axisymmetric solution with $J_{AB} = 0$ as follows. The most general diagonal traceless and divergence free stationary $T^{ab}$ is given by 
\begin{equation}
\begin{split}
&T^{tt}_{\text{corr}} = \ell^2 [2 T^{\theta\theta}(\theta) + \tan \theta ~{T^{\theta \theta}}^\prime  (\theta)], \quad T^{\theta \theta}_{\text{corr}}  = T^{\theta \theta}(\theta), \\
&T^{\phi\phi}_{\text{corr}}  = \frac{1}{\sin^2 (\theta)} [T^{\theta\theta}(\theta) + \tan \theta ~{T^{\theta\theta}}^\prime(\theta) ]
\end{split}
\end{equation} and the other components are set to zero. We consider the sum of $T_{\text{Kerr}}+T_{\text{corr}}$. We solve for $T^{\theta\theta}(\theta)$ to set $J^{AB} = 0$. The regular solution at $\mathscr I$ is unique and given by
\begin{eqnarray}
T^{tt}= -\frac{m \ell^3}{4\pi},\qquad T^{t\phi} = -\frac{3a m \ell \gamma^5}{8 \pi }, \qquad T^{AB} = -\frac{m \ell}{8 \pi} q^{AB}. 
\end{eqnarray}
The mass and angular momentum are $M  = \int \sqrt{-g_{(0)}} \, T^t_{\,\, t} = m$,  $J= -\int \sqrt{-g_{(0)}} \, T^t_{\,\,\phi} = \frac{ma}{\Xi^2}$. It would be interesting to know whether this solution is regular in the bulk of spacetime. 

From the conservation of the stress-energy tensor $T^{ab}$ given by the first equation of \eqref{condition on energy momentum}, the most general stationary solution with flat boundary metric \eqref{flat} is only constrained by the following conditions:
\begin{equation}
D_A N^A = 0 \Leftrightarrow N^A = \epsilon^{AB} D_B \alpha (x^C), \quad \partial_A M = 0.\label{soll}
\end{equation} $\alpha (x^C)$ is an arbitrary function of $x^C$. To obtain these expressions, we also used equations \eqref{eq:RefiningTab} and \eqref{BCads}. Therefore, even for stationary solutions, we see that the charges associated with the area-preserving diffeomorphisms are generically non-vanishing. It would be interesting to study the regularity of the general solutions \eqref{soll} in the bulk of spacetime.

\section*{Acknowledgments}
We thank Yegor Korovin for useful discussions. We gratefully thank Marios Petropoulos, David Rivera-Betancour and Amitabh Virmani for pointing out typos in the second version of this manuscript which led to a Corrigendum, and Jahanur Hoque for spotting one further typo in Eq. (2.25). G.C., A.F. and R.R. are, respectively, Research Associate, Research Fellow and FRIA Research Fellow of the Fonds de la Recherche Scientifique F.R.S.-FNRS (Belgium).

\appendix

\section{Killing vectors of AdS$_\mathbf{4}$}
\label{app:ExactVectors}

AdS$_4$ is isometrically immersed into $\mathbb{R}^{(2,3)}$ as the hypersurface
\begin{equation}
\lbrace X^\mu \in \mathbb{R}^{(2,3)} | -X_0^2 - X_{\bar{0}}^2 + X_1^2 +X_2^2+X_3^2 = -\ell^2 \rbrace .
\end{equation}
The symmetry group of AdS$_4$ is the homogeneous part of $ISO(2,3)$, the isometry group of $\mathbb{R}^{(2,3)}$, which is $SO(2,3)$. The generators of $SO(2,3)$ algebra 
\begin{equation}
\mathcal J_{ab} = \mathcal J_{[ab]} = X_b \partial_a - X_a \partial_b
\end{equation}
directly lead to the Killing vectors of AdS$_4$ after pullback. 

In retarded coordinates $(u,r,x^A)$, the AdS$_4$ line element takes the form 
\begin{equation}
ds^2 = -\Big(\frac{r^2}{\ell^2}+ 1 \Big)du^2 - 2dudr +r^2 \mathring q_{AB} dx^A dx^B,\label{AdSBondi}
\end{equation}
where we take $x^A=(z,\bar z)$ as the stereographic coordinates on the two sphere of unit round metric $\mathring q_{AB}$. The Minkowski metric in retarded coordinates is recovered in the flat limit $\ell \rightarrow \infty$. In these coordinates, the Killing vectors are given by 
\begin{eqnarray}
\mathcal J_{ab}^u = f, \qquad \mathcal J_{ab}^r = -\frac{r}{2} D_A \mathcal J_{ab}^A,\qquad \mathcal J_{ab}^A = Y^A - \frac{1}{r} \partial^A f 
\end{eqnarray}
where $D_A$ is the covariant derivative on the sphere and the functions $f, Y^A$ are constrained as 
\begin{equation}
\partial_u f = \frac{1}{2} D_A  Y^A \, , \qquad
\partial_u Y^A = \frac{1}{\ell^2} \partial^A f \, , \qquad 2 D_{(A} Y_{B)} - q_{AB} D_C Y^C = 0. 
\end{equation}
Explicitly, the Killing vectors are given by
\begin{enumerate}
\item Rotations:
\begin{equation}
\begin{split}
\mathcal J_{12} &= [0,0,-i z,-i \bar{z}] = \partial_\phi, \\
\mathcal J_{13} &= [0,0,\frac{1}{2} (-1-z^2),\frac{1}{2} (-1-\bar{z}^2)], \\
\mathcal J_{23} &= [0,0,\frac{1}{2} i (-1 + z^2),\frac{1}{2} i (-1 + \bar{z}^2)];
\end{split}
\end{equation}

\item Time translation: $ \mathcal J_{0\bar{0}} = \partial_u$;

\item Boosts on the first timelike coordinate: with the shorthand notation $M(\frac{u}{\ell},r) \equiv r \cos (u/\ell) + \ell \sin (u/\ell)$, we have
\begin{equation}
\begin{split}
\mathcal J_{01} &= \Big[ \frac{(z+\bar{z}) \ell \sin (\frac{u}{\ell})}{1+z\bar{z}}, -\frac{(z+\bar z)M(\frac{u}{\ell},r)}{(1+z\bar{z})},\frac{(-1+z^2) M(\frac{u}{\ell},r)}{2 r},-\frac{(-1+\bar{z}^2) M(\frac{u}{\ell},r)}{2 r}  \Big], \\
\mathcal J_{02} &= \Big[ -\frac{(z-\bar{z}) \ell \sin (\frac{u}{\ell})}{1+z\bar{z}}, \frac{i(z-\bar z)M(\frac{u}{\ell},r)}{(1+z\bar{z})},-\frac{i(1+z^2) M(\frac{u}{\ell},r)}{2 r},\frac{i(1+\bar{z}^2) M(\frac{u}{\ell},r)}{2 r }  \Big], \\
\mathcal J_{03} &= \Big[ \frac{(-1+z\bar{z}) \ell \sin (\frac{u}{\ell})}{1+z\bar{z}}, \frac{(-1+z\bar z)M(\frac{u}{\ell},r)}{(1+z\bar{z})},-\frac{z M(\frac{u}{\ell},r)}{r},-\frac{\bar z M(\frac{u}{\ell},r)}{r}  \Big] ;\\
\end{split}
\end{equation}

\item Boosts on the second timelike coordinate: by virtue of the $SO(2,3)$ algebra,
\begin{equation}
\mathcal{J}_{\bar 0 i} = -[\mathcal{J}_{0\bar 0},\mathcal{J}_{0i}] = - \partial_u \mathcal{J}_{0i}.
\end{equation}

\end{enumerate}
The 4 first vectors are Killing symmetries of the cylindric boundary (manifest in Fefferman-Graham gauge), and the 6 other vectors are conformal Killing symmetries of that same boundary.

The Lorentz algebra can be explicitly checked for the rotations $V_i = \epsilon_{ijk} J_{jk}$ together with either set of  boosts $K_i = \mathcal  J_{0i}$ or $\ell \mathcal J_{\bar{0}i}$:
\begin{equation}
[ V_i,V_j] = \epsilon_{ijk} V_k \, , \quad [ K_i,K_j] = -\epsilon_{ijk} V_k \, , \quad [V_i,K_j] = \epsilon_{ijk} K_k.
\end{equation}

In the flat limit, time translations and rotations are trivial. The boosts $\mathcal J_{0i} $ become the Lorentz boosts 
\begin{equation}
\begin{split}
\mathcal J_{01} &\rightarrow \Big[  \frac{u(z+\bar z)}{1+z\bar z}, -\frac{(r+u)(z+\bar{z})}{1+z\bar{z}} , \frac{(r+u)(-1+z^2)}{2 r} , \frac{(r+u)(-1+\bar{z}^2)}{2 r}  \Big], \\
\mathcal J_{02} &\rightarrow \Big[  -i\frac{u(z-\bar z)}{1+z\bar z}, i\frac{(r+u)(z-\bar{z})}{1+z\bar{z}} , -i\frac{(r+u)(1+z^2)}{2 r} , i\frac{(r+u)(1+\bar{z}^2)}{2 r}  \Big], \\
\mathcal J_{03} &\rightarrow \Big[  \frac{u(-1+z\bar z)}{1+z\bar z}, -\frac{(r+u)(-1+z\bar{z})}{1+z\bar{z}} , -\frac{(r+u)z}{r} , -\frac{(r+u)\bar{z}}{r}  \Big],
\end{split}
\end{equation}
and the boosts $\mathcal J_{\bar 0i} $ become the spatial translations
\begin{equation}
\begin{split}
\mathcal J_{\bar 01} &\rightarrow \Big[ \frac{z+\bar{z}}{1+z\bar{z}} , - \frac{z+\bar{z}}{1+z\bar{z}} , \frac{-1+z^2}{2r} , \frac{-1+\bar{z}^2}{2r} \Big], \\
\mathcal J_{\bar 02} &\rightarrow \Big[ -i \frac{z-\bar{z}}{1+z\bar{z}} , i \frac{z-\bar{z}}{1+z\bar{z}} , -i \frac{1+z^2}{2r}, i \frac{1+\bar{z}^2}{2r} \Big], \\
\mathcal J_{\bar 03} &\rightarrow \Big[ - \frac{1-z\bar{z}}{1+z\bar{z}} , \frac{1-z\bar{z}}{1+z\bar{z}} , -\frac{z}{r}, -\frac{\bar{z}}{r} \Big].
\end{split}
\end{equation}

\section{Map from Bondi to Fefferman-Graham gauge}
\label{app:chgt}

In this section, we find the explicit change of coordinates that maps a general vacuum asymptotically locally (A)dS$_4$ spacetime ($\Lambda \neq 0$) in Bondi gauge to Fefferman-Graham gauge. This procedure will lead to the  explicitly map between the free functions defined in Bondi gauge $ \lbrace q_{AB}, \beta_0,U^A_0,\mathcal{E}_{AB},M,N_A\rbrace $ and the holographic functions defined in Fefferman-Graham gauge, namely the boundary metric $g_{ab}^{(0)}$ and the boundary stress-tensor $g^{(3)}_{ab}$. 

We follow and further develop the procedure introduced in \cite{Poole:2018koa}. We first note that one can map the AdS$_4$ vacuum metric in retarded coordinates \eqref{AdSBondi} to the global patch 
\begin{equation}
ds^2 = -\Big(\frac{r^2}{\ell^2}+ 1 \Big)dt^2 + \Big(\frac{r^2}{\ell^2}+ 1 \Big)^{-1} dr^2 +r^2 \mathring q_{AB} dx^A dx^B
\end{equation}
by using $u = t - r_\star$ where the tortoise coordinate is $r_\star \equiv \ell [\arctan \left( \frac{r}{\ell} \right)  -\frac{\pi}{2}]$, which maps $r=\infty$ to $r_\star=0$. The change of coordinates from $(t,r_\star,x^A)$ to Fefferman-Graham gauge $(t,\rho,x^A)$ can then be performed perturbatively in series of $\rho$ around $\rho = 0$, identified with $r_\star = 0$. 

The general algorithm is then the following:
\begin{enumerate}
\item Starting from any asymptotically locally AdS$_4$ solution formulated in Bondi gauge $(u,r,x^A)$, we perform the preliminary change to the tortoise radial coordinate,
\begin{equation}
\begin{split}
u &\to t-r_\star, \quad x^A \to x^A, \\
r &\to \ell \tan \Big[\frac{r_\star}{\ell}+\frac{\pi}{2}\Big] = -\frac{\ell^2}{r_\star} + \frac{r_\star}{3} + \frac{r_\star^3}{45\ell^2} + \mathcal{O}(r_\star^{5}).
\end{split}
\end{equation}
\item We reach the Fefferman-Graham gauge at order $N \geq 0$ perturbatively,
\begin{equation}
g_{\rho\rho} = -\frac{3}{\Lambda}\frac{1}{\rho^2} \Big( 1 + \mathcal{O}(\rho^{N+1}) \Big), \quad g_{\rho t} = \frac{1}{\rho^2}\mathcal{O}(\rho^{N+1}), \quad g_{\rho A} = \frac{1}{\rho^2}\mathcal{O}(\rho^{N+1}), \label{eq:FGgaugecond}
\end{equation} thanks to a second change of coordinates,
\begin{equation}
\begin{split}
r_\star &\to \sum_{n=1}^{N+1} R_n (t,x^A)\rho^n, \\
t &\to t + \sum_{n=1}^{N+1} T_n (t,x^A)\rho^n, \\
x^A &\to x^A + \sum_{n=1}^{N+1} X^A_n (t,x^B)\rho^n.
\end{split}
\end{equation}
\end{enumerate}
In order to obtain all the free functions in $\gamma_{ab}$, we need to proceed up to order $N=3$. For each $n$, each gauge condition \eqref{eq:FGgaugecond} can be solved separately and will determine algebraically $R_n$, $T_n$ and $X_n^A$ respectively. Only the function $R_1(t,x^A)$ remains unconstrained by these conditions, since it represents a Weyl transformation on the boundary metric that is allowed within Fefferman-Graham gauge. This Weyl transformation is constrained by the choice of luminosity distance $r$ in Bondi coordinates which ensures $g_{AB}^{(0)}=q_{AB}$.

We use the following shorthand notations for subleading fields in Bondi gauge:
\begin{equation}
\begin{split}
\frac{V}{r} &= \frac{\Lambda}{3}e^{2\beta_0}r^2 + r \ V_{(1)} (t,x^A) + V_{(0)} (t,x^A) + \frac{2M}{r} + \mathcal{O}(r^{-2}), \\
U^A &= U_{0}^A(t,x^B) + \frac{1}{r}U_{(1)}^A(t,x^B) + \frac{1}{r^2} U_{(2)}^A(t,x^B) + \frac{1}{r^3} U_{(3)}^A(t,x^B) + \mathcal{O}(r^{-4}), \\
\beta &= \beta_{0}(t,x^A) + \frac{1}{r^2}\beta_{(2)}(t,x^A)+ \mathcal{O}(r^{-4}).
\end{split}
\end{equation}
whose explicit on-shell values can be read off in \eqref{eq:EOMVr} and \eqref{eq:EOM_UA2}. That will state the equations in a more compact way. All the fields are now evaluated on $(t,x^A)$ since the time coordinate on the boundary can be defined as $t$ as well as $u$. We also define  some recurrent structures appearing in the diffeomorphism as differential operators on boundary scalar fields $f(t,x^A)$:
\begin{equation}
\begin{split}
P[f] &= \frac{1}{2} e^{-4\beta_0} (\partial_t f + U_0^A \partial_A f), \\
Q[f;g] &= P[f] - 2  P[g] f, \\
B_A[f] &= \frac{1}{2} e^{-2\beta_0} (\partial_A - 2\partial_A \beta_0) f.
\end{split}
\end{equation}
$P^n[f]$ denotes $n$ applications of $P$ on $f$, for example $P^2[f] \equiv P[P[f]]$. Now we can write down the perturbative change of coordinate to Fefferman-Graham gauge:

\begin{align*}
R_1 (t,x^A) &= -\frac{3}{\Lambda}, \\
R_2 (t,x^A) &= \frac{9}{2\Lambda^2} e^{-2\beta_0} V_{(1)}, \\
R_3 (t,x^A) &= \frac{3}{2\Lambda} \beta_{(2)} - \frac{3}{\Lambda^2} \Big( 1 + \frac{3}{4} e^{-2\beta_0} V_{(0)} \Big) +\frac{27}{2\Lambda^3}  \Big( Q[V_{(1)};\beta_0] - \frac{3}{8} e^{-4\beta_0} V_{(1)}^2 \Big), \\
R_4 (t,x^A) &= \frac{3}{\Lambda^2} e^{-2\beta_0} \Big( M + 2 e^{4\beta_0} P[\beta_{(2)}] - \frac{5}{2} V_{(1)} \beta_{(2)}  \Big) \\
	&\hspace{-40pt}\quad -\frac{9}{\Lambda^3} \Big\lbrace Q[V_{(0)};\beta_0] + \frac{1}{4} e^{-4\beta_0} \Big[ U^A_{(1)} \partial_A V_{(1)} - 2 V_{(1)} U^A_{(1)} \partial_A\beta_0 - 3 V_{(1)} (2 e^{2\beta_0} + V_{(0)}) \Big] \Big\rbrace \\
	&\hspace{-40pt}\quad + \frac{27}{\Lambda^4} e^{2\beta_0} \Big[ P^2[V_{(1)}] - 2 V_{(1)} \Big( P^2[\beta_0]+ \frac{1}{2}e^{-4\beta_0} Q[V_{(1)};\beta_0] - \frac{3}{32} e^{-8\beta_0} V_{(1)}^2 \Big) - 2 P[\beta_0]P[V_{(1)}] \Big], \\
	&\ \\
T_1 (t,x^A) &= (1 - e^{-2\beta_0}) R_1(t,x^A),\\
T_2 (t,x^A) &=  (1 - e^{-2\beta_0}) R_2 (t,x^A) -\frac{18}{\Lambda^2} \Big( P[\beta_0] - \frac{1}{4}e^{-4\beta_0} V_{(1)} \Big),\\
T_3 (t,x^A) &=  (1- e^{-2\beta_0})R_3(t, x^A) -\frac{3}{\Lambda^2} e^{-2\beta_0} (1 + e^{-2\beta_0} V_{(0)} - 2 \partial^A \beta_0 \partial_A \beta_0) \\
	&\quad + \frac{9}{\Lambda^3} e^{-2\beta_0} \Big( Q[V_{(1)};		\beta_0] - 4 e^{4\beta_0} P^2[\beta_0]- \frac{1}{2}		e^{-4\beta_0} V_{(1)}^2 \Big), \\
T_4 (t,x^A) &= (1- e^{-2\beta_0})R_4(t, x^A) \\
& \quad+ \frac{9}{2\Lambda^2} \Big[ e^{-4\beta_0} \Big( M - \beta_{(2)} V_{(1)} - \frac{1}{3} U_{(2)}^A \partial_A \beta_0 \Big) -\frac{1}{2} (P[\beta_{(2)}] - 8 \beta_{(2)} P[\beta_0]) \Big] \\
	&\quad - \frac{27}{\Lambda^3} \Big\lbrace \frac{1}{8} e^{-2\beta_0} \Big( 3 Q[V_{(0)};\beta_0] - \frac{8}{3} P[\beta_0]V_{(0)} - 2 e^{-4\beta_0} V_{(1)}V_{(0)} \Big) \\
		&\qquad\qquad + \frac{1}{3} e^{-2\beta_0} \Big( P[U_{(1)}^A]\partial_A\beta_0 + \frac{3}{2} U_{(1)}^A \partial_A P[\beta_0] \Big) \\
		&\qquad\qquad - \frac{1}{12}e^{-4\beta_0} \Big[ U_{(1)}^A B_A [V_{(1)}] + 6 V_{(1)} - 2 (V_{(1)}\partial_A \beta_0 + 2 \partial_B \beta_0 \partial_A U_0^B)\partial^A \beta_0 \Big] \Big\rbrace \\
	&\quad + \frac{81}{\Lambda^4} \Big\lbrace -\frac{2}{3} e^{4\beta_0} \Big( P^3[\beta_0] + 2 P[\beta_0]P^2[\beta_0] \Big) + \frac{1}{4} \Big(P^2[V_{(1)}] - 2 V_{(1)} P^2[\beta_0]\Big) \\
	&\qquad\qquad +\frac{1}{6} P[\beta_0] \Big[\Big(\frac{13}{4}e^{-4\beta_0} V_{(1)} - 8 P[\beta_0] \Big) V_{(1)} + P[V_{(1)}] \Big] \\
	&\qquad\qquad - \frac{1}{16} e^{-4\beta_0} V_{(1)} \Big(5P[V_{(1)}]- e^{-4\beta_0} V_{(1)}^2\Big) \Big\rbrace, \\
&\ \\
X_1^A (t,x^B) &= (T_1 - R_1) U_0^A,\\
X_2^A (t,x^B) &= (T_2 - R_2) U_0^A - \frac{3}{2\Lambda} e^{-2\beta_0} U_{(1)}^A + \frac{9}{\Lambda^2} P[U_0^A],\\
X_3^A (t,x^B) &= (T_3 - R_3) U_0^A + \frac{1}{\Lambda} e^{-2\beta_0} U_{(2)}^A \\
	&\quad - \frac{6}{\Lambda^2} \Big[ Q[U_{(1)}^A;\beta_0] + \frac{1}{2} B^A[V_{(1)}] + \frac{1}{4} e^{-4\beta_0} (U_{(1)}^B \partial_B U^A_0 - V_{(1)}U_{(1)}^A) \Big] \\
		&\quad + \frac{18}{\Lambda^3} e^{2\beta_0} Q[P[U_0^A];\beta_0], \\
X_4^A (t,x^B) &= (T_4 - R_4) U_0^A - \frac{3}{4\Lambda} e^{-2\beta_0} \Big[ U_{(3)}^A + \frac{1}{2}e^{2\beta_0} (\partial^A \beta_{(2)} - 8 \beta_{(2)}\partial^A\beta_0) \Big] \\
&\quad + \frac{9}{2\Lambda^2} \Big[ Q[U_{(2)}^A;\beta_0] - \frac{1}{2} e^{-4\beta_0} \Big( V_{(1)} U_{(2)}^A - \frac{1}{3} U_{(2)}^B \partial_B U_{(0)}^A \Big) - 2 \beta_{(2)} P[U_0^A] \\
		&\qquad\qquad + \frac{1}{4} B^A[V_{(0)}] + \frac{1}{2} C^{AC} B_C[V_{(1)}] + \frac{1}{2} e^{-2\beta_0} U_{(1)}^C B_C[ U^A_{(1)} ] \Big] \\
		&\quad -\frac{27}{\Lambda^3} \Big\lbrace  e^{2\beta_0} \Big( P[Q[U_{(1)}^A;\beta_0]] + P[B^A[V_{(1)}]] - \frac{1}{2} q^{AC} P[B_C[V_{(1)}]] \Big) \\
		&\qquad\qquad - \frac{1}{2}e^{-2\beta_0} \Big[ V_{(1)} P[U_{(1)}^A] - \frac{2}{3}\Big( P[U_{(1)}^C] + \frac{1}{2} B^C[V_{(1)}] - 5 P[\beta_0]U_{(1)}^C \Big)\partial_C U_0^A \\
		&\qquad\qquad + \frac{1}{2} P[V_{(1)}]U_{(1)}^A - \frac{2}{3}(V_{(0)} - 8 e^{2\beta_0} \partial^B \beta_0 \partial_B \beta_0)P[U_0^A] - U_{(1)}^B P[\partial_B U_0^A] \\
		&\qquad\qquad + \frac{1}{2} (\partial^A U_0^C)B_C[V_{(1)}] \Big] + 3 P[\beta_0] V_{(1)}\partial^A \beta_0 \\
		&\qquad\qquad - e^{-4\beta_0} \Big[ \frac{3}{32} \Big(\partial^A (V_{(1)}^2) - \frac{20}{3} \partial^A \beta_0 V_{(1)}^2 \Big) + \frac{1}{6} (V_{(1)}  \partial^B \beta_0 - \partial^C\beta_0 \partial_C U_0^B) \partial_B U_0^A \Big] \Big\rbrace. \\ 
		&\quad +\frac{81}{\Lambda^4} \Big\lbrace \frac{1}{3} e^{4\beta_0} P^3[U_0^A] + \Big[ \frac{1}{4}e^{-4\beta_0} V_{(1)}^2 - \frac{1}{3} Q[V_{(1)};\beta_0] - \frac{4}{3} e^{4\beta_0} (P^2[\beta_0] + P[\beta_0]^2)  \Big] P[U_0^A] \Big\rbrace. \nonumber
\end{align*}

Several consistency checks can be performed at each stage of the computation. The boundary metric in Fefferman-Graham gauge must be equivalent to the pulled-back metric on the hypersurface $\lbrace r\to \infty \rbrace$ in Bondi gauge, up to the usual replacement $u\to t$:
\begin{equation}
g_{ab}^{(0)} = \left[
\begin{array}{cc}
\frac{\Lambda}{3} e^{4\beta_0} + U_0^C U^0_C & -U^0_B \\ 
-U^0_A & q_{AB}
\end{array} 
\right].
\end{equation}
At subleading orders, $g^{(1)}_{ab}$ and $g^{(2)}_{ab}$ must be algebraically determined by $g_{ab}^{(0)}$ and its first and second derivatives, what it turns out to be the case. The constraint \eqref{eq:CAB} forces $g^{(1)}_{ab} = 0$ while the annulation of $\mathcal{D}_{AB}(t,x^C)$ \eqref{eq:DAB} results in
\begin{equation}
g^{(2)}_{ab} = \frac{3}{\Lambda} \Big[ R^{(0)}_{ab} - \frac{1}{4} R_{(0)} g^{(0)}_{ab} \Big].
\end{equation}
We will not give the full general form of $g^{(3)}_{ab}$, but it can be proven that this tensor is traceless with respect to $g^{(0)}_{ab}$, and that the equations of motion in Bondi gauge are necessary and sufficient to show its conservation $D^{(0)}_a g_{(3)}^{ab} = 0$, as we argued in the main text.

After boundary gauge fixing $\beta_0 = 0$, $U_0^A = 0$, the expressions of each coefficient in the diffeomorphism simplify drastically:
\begin{align*}
R_1 (t,x^A) &= -\frac{3}{\Lambda}, \\
R_2 (t,x^A) &= \frac{9}{2\Lambda^2} V_{(1)}, \\
R_3 (t,x^A) &= \frac{3}{2\Lambda} \beta_{(2)} - \frac{3}{\Lambda^2} \Big( 1 + \frac{3}{4} V_{(0)} \Big) +\frac{27}{2\Lambda^3}  \Big( \frac{1}{2} \partial_t V_{(1)} - \frac{3}{8} V_{(1)}^2 \Big), \\
R_4 (t,x^A) &= \frac{3}{\Lambda^2} \Big( M + \partial_t \beta_{(2)} - \frac{5}{2} V_{(1)} \beta_{(2)} \Big) -\frac{9}{\Lambda^3} \Big[ \frac{1}{2} \partial_t V_{(0)} - \frac{3}{4} V_{(1)} (2 + V_{(0)}) \Big] \\
	&\quad + \frac{27}{\Lambda^4} \Big( \frac{1}{4}\partial_t^2[V_{(1)}] - \frac{1}{4}\partial_t V_{(1)}^2 + \frac{6}{32} V_{(1)}^3\Big). \\
	&\ \\
T_1 (t,x^A) &= 0,\\
T_2 (t,x^A) &=  \frac{9}{2\Lambda^2} V_{(1)},\\
T_3 (t,x^A) &=  -\frac{3}{\Lambda^2} (1 + V_{(0)}) + \frac{9}{2\Lambda^3} \Big( \partial_t V_{(1)} - V_{(1)}^2 \Big), \\
T_4 (t,x^A) &= \frac{9}{2\Lambda^2} \Big[ M - \frac{1}{4} (\partial_t \beta_{(2)} + 4 V_{(1)}\beta_{(2)}) \Big] + \frac{9}{\Lambda^3} \Big[ - \frac{9}{16} \partial_t V_{(0)} + \frac{3}{2} V_{(1)} (1+\frac{1}{2} V_{(0)}) \Big] \\
	&\quad + \frac{81}{\Lambda^4} \Big(  \frac{1}{16} \partial_t^2 V_{(1)}  - \frac{5}{64} \partial_t V_{(1)}^2 + \frac{1}{16}V_{(1)}^3 \Big). \\
	&\ \\	
X_1^A (t,x^B) &= X_2^A (t,x^B) = 0, \\
X_3^A (t,x^B) &= \frac{1}{\Lambda} U_{(2)}^A - \frac{3}{2\Lambda^2} \partial^A V_{(1)}, \\
X_4^A (t,x^B) &= - \frac{3}{4\Lambda} \Big( U_{(3)}^A + \frac{1}{2}\partial^A \beta_{(2)} \Big) + \frac{9}{2\Lambda^2} \Big( \frac{1}{2}\partial_t U_{(2)}^A - \frac{1}{2}  V_{(1)} U_{(2)}^A + \frac{1}{8} \partial^A V_{(0)} \Big) \\
		&\quad -\frac{27}{16\Lambda^3}  \ q^{AB} \Big( \partial_t \partial_B V_{(1)} + \frac{1}{2} V_{(1)} \partial_B V_{(1)} \Big).
\end{align*}

The expressions of the diffeomorphism and the metric elements $g_{ab}^{(0)}$, $g_{ab}^{(2)}$ and $g_{ab}^{(3)}$ (the latter within the additional boundary gauge fixing) can be found in the attached \texttt{Mathematica} file, \texttt{BMS\_TO\_FG.nb}. Please note that the package \texttt{RGtensors} is needed to run this notebook.

\newpage

\providecommand{\href}[2]{#2}\begingroup\raggedright\endgroup

\end{document}